\documentclass[twocolumn,astrosymb,tighten,times]{aastex631}

\newcommand\aastex{AAS\TeX}
\newcommand\latex{La\TeX}

\received{November 29, 2023}
\revised{April 28, 2025}
\accepted{May 25, 2025}

\submitjournal{Astrophysical Journal}

\usepackage{amsmath}
\usepackage{mathrsfs}
\usepackage{graphicx}
\usepackage{dcolumn}
\usepackage{multirow}
\usepackage{bm}
\usepackage{textcomp}
\usepackage{longtable}
\usepackage{booktabs}
\usepackage{epsfig}
\usepackage{rotating}
\usepackage{xfrac}
\usepackage{threeparttable}
\usepackage{lipsum} 
\usepackage{float}
\RequirePackage{hyperref}
\hypersetup
{
unicode=false,%
pdftoolbar=true,
pdfmenubar=true,
pdffitwindow=false,
pdfstartview=FitH,
pdftitle={ApJ},
pdfauthor={LAM Yi Hua},
pdfsubject={Nuclear Astrophysics},
pdfcreator={LAM Yi Hua},
pdfproducer={LAM Yi Hua},
pdfkeywords={NucAstro},
bookmarksnumbered,%
colorlinks=true,%
linkcolor=magenta,%
citecolor=blue,
filecolor=cyan,%
urlcolor=blue,
breaklinks, 
plainpages=false}
\usepackage{CJKutf8}
\newcommand{\cntextSimKai}[1]{\begin{CJK*}{UTF8}{gkai}#1\end{CJK*}}
\newcommand{\cntextTraKai}[1]{\begin{CJK*}{UTF8}{bkai}#1\end{CJK*}}
\newcommand{\jptextJap}[1]{\begin{CJK*}{UTF8}{min}#1\end{CJK*}}
\usepackage{xcolor}
\definecolor{Mycolor}{HTML}{C5E0B3}
\definecolor{RED}{HTML}{ee6677}
\definecolor{red}{HTML}{ee6677}
\definecolor{blue}{HTML}{4477aa}
\definecolor{green}{HTML}{228833}
\definecolor{magenta}{HTML}{ee3377}
\definecolor{cyan}{HTML}{66ccee}
\definecolor{yellow}{HTML}{ccbb44}
\definecolor{grey}{HTML}{bbbbbb}
\shortauthors{L\lowercase{am et al}.}

\begin{document}

\title{The Regulated GeAs Cycles with the New $^{63}$Ga(p,$\gamma$)$^{64}$Ge and $^{64}$Ge(p,$\gamma$)$^{65}$As Reaction Rates and Its Impact on the GS~1826$-$24 Clocked Bursts and SAX~J1808.4$-$3658 Photospheric Radius Expansion Bursts}

\correspondingauthor{Y. H. Lam, A. Heger}
\email{lamyihua@zstu.edu.cn, alexander.heger@monash.edu}

\author[0000-0001-6646-0745]{Yi Hua Lam (\cntextTraKai{藍乙華})}
\affiliation{Zhejiang Key Laboratory of Quantum State Control and Optical Field Manipulation, \\Department of Physics, \href{https://ror.org/03893we55}{Zhejiang Sci-Tech University}, Hangzhou 310018, China; \href{mailto:lamyihua@zstu.edu.cn}{lamyihua@zstu.edu.cn}}
\affiliation{Astrophysical Big Bang Laboratory, Cluster for Pioneering Research, \href{https://ror.org/01sjwvz98}{RIKEN}, Wako, Saitama 351-0198, Japan}
\affiliation{\href{https://ror.org/03x8rhq63}{Institute of Modern Physics}, Chinese Academy of Sciences, Lanzhou 730000, China}

\author[0000-0002-3445-0451]{Ning Lu (\cntextTraKai{盧寧})}
\affiliation{School of Nuclear Science and Technology, \href{https://ror.org/01mkqqe32}{Lanzhou University}, Lanzhou 730000, China}
\affiliation{\href{https://ror.org/03x8rhq63}{Institute of Modern Physics}, Chinese Academy of Sciences, Lanzhou 730000, China}


\author[0000-0002-3684-1325]{Alexander Heger}
\affiliation{School of Physics and Astronomy, \href{https://ror.org/02bfwt286}{Monash University}, Vic 3800, Australia; \href{mailto:alexander.heger@monash.edu}{alexander.heger@monash.edu}} 
\affiliation{Argelander-Institute for Astronomy, \href{https://ror.org/041nas322}{University of Bonn}, Auf dem H{\"u}gel 71, Bonn 53121, Germany}
\affiliation{Joint Institute for Nuclear Astrophysics, \href{https://ror.org/05hs6h993}{Michigan State University}, East Lansing, MI 48824, USA}

\author[0000-0001-5652-1516]{Zi Xin Liu (\cntextSimKai{刘子鑫})}
\affiliation{Advanced Energy Science and Technology, Guangdong Laboratory, Huizhou 516000, China}
\affiliation{\href{https://ror.org/03x8rhq63}{Institute of Modern Physics}, Chinese Academy of Sciences, Lanzhou 730000, China}

\author[0000-0003-4023-4488]{Zac~Johnston}
\affiliation{Joint Institute for Nuclear Astrophysics, \href{https://ror.org/05hs6h993}{Michigan State University}, East Lansing, MI 48824, USA} 
\affiliation{Department of Physics and Astronomy, \href{https://ror.org/05hs6h993}{Michigan State University}, East Lansing, MI 48824, USA}

\author[0000-0001-9931-2052]{Hidetoshi~Yamaguchi (\jptextJap{山口英斉})}
\affiliation{Center for Nuclear Study, \href{https://ror.org/057zh3y96}{the University of Tokyo}, RIKEN campus, 2-1 Hirosawa, Wako, Saitama 351-0198, Japan} 
\affiliation{\href{https://ror.org/052rrw050}{National Astronomical Observatory of Japan}, 2-21-1 Osawa, Mitaka, Tokyo 181-8588, Japan}


\begin{abstract}
\footnotesize
The $^{63}$Ga(p,$\gamma$)$^{64}$Ge and $^{64}$Ge(p,$\gamma$)$^{65}$\textcolor{black}{As} \textcolor{black}{thermonuclear reactions connect the ZnGa and GeAs cycles by diverting the flow of the rapid proton-capture process from $^{63}$Ga to $^{65}$As. Changes in these two reaction rates regulate the ZnGa and GeAs cycles and may affect the modeled properties matching with the observed counterparts of a type I X-ray burster. 
We implement the latest $^{63}$Ga(p,$\gamma$)$^{64}$Ge and $^{64}$Ge(p,$\gamma$)$^{65}$As reaction rates to the state-of-the-art self-consistent one-dimensional multizone thermohydrodynamic code, \textsc{Kepler}, to study the influence of these new reaction rates on the models of the GS~1826$-$24 clocked burster and SAX~J1808.4$-$3658 photospheric radius expansion burster. 
Both new reaction rates obtained by \href{https://doi.org/10.1103/PhysRevC.110.065804}{Lu \emph{et al.} (2024)} are determined from complementing the experimental input with the nuclear spectroscopic information deduced from the full \emph{pf}-shell space configuration interaction shell model calculations.} 
By constraining the models on reproducing the observed burst peak, light-curve profile, fluence, and recurrence time, we find that the impact of the newly measured proton thresholds and respective proton-capture reactions on the burst light-curve profile of the GS~1826$-$24 clocked burster is\textcolor{black}{, in fact,} not as significant as claimed by Zhou \emph{et al}.~[\href{https://doi.org/10.1038/s41567-023-02034-2}{Nat. Phys. 19, 1091 (2023)}]. \textcolor{black}{With or without the inclusion of the newly determined reaction rate of the highly influential $^{22}$Mg($\alpha$,p)$^{25}$Al reaction, the impact of the new $^{63}$Ga(p,$\gamma$)$^{64}$Ge and $^{64}$Ge(p,$\gamma$)$^{65}$As reaction rates} on SAX~J1808.4$-$3658 photospheric radius expansion bursts is evident. \textcolor{black}{Our finding indicates that the models reproducing the 2002 October epoch of the SAX~J1808.4$-$3658 photospheric radius expansion burster are more sensitive to the uncertainties of thermonuclear reaction rates.} 
\end{abstract}

\keywords{\footnotesize 
\href{http://astrothesaurus.org/uat/2081}{Reaction rates (2081)}; 
\href{http://astrothesaurus.org/uat/1965}{Computational methods (1965)}; 
\href{http://astrothesaurus.org/uat/1108}{Neutron stars (1108)}; 
\href{http://astrothesaurus.org/uat/939}{Low-mass X-ray binary stars (939)}; 
\href{http://astrothesaurus.org/uat/1814}{X-ray bursts (1814)}; 
\href{http://astrothesaurus.org/uat/1813}{X-ray bursters (1813)}; 
\href{http://astrothesaurus.org/uat/1811}{X-ray binary stars (1811)};
\href{http://astrothesaurus.org/uat/739}{High energy astrophysics (739)}
}


\section{Introduction} \label{sec:intro}

Key alpha-proton ($\alpha$p-) and rapid-proton-capture (rp-) processes play the important role of determining the energy release and final isotopic abundance of burst ash in thermonuclear (type I) X-ray bursts (XRBs; \citealt{Wallace1981, Woosley2004, Fisker2008, Jose2010}). These key thermonuclear reactions of proton-rich nuclei coupled with $\beta$ decays, astrophysical conditions, and thermohydrodynamics determine the main reaction flow of the commonly designated rp-process path near the proton drip line and the incorporated nucleosynthesis during the thermonuclear runaway of type I XRBs that happens in the accreted envelope above the surface of the host neutron star of a low-mass X-ray binary. 

The pioneer study of modeling the GS~1826$-$24 clocked bursts, based on the self-consistent one-dimensional multizone thermohydrodynamic \textsc{Kepler} code, provides a gross picture of the interplay between thermohydrodynamics, nucleosynthesis, and nuclear energy generation during the evolution of an accreted envelope igniting XRBs \citep{Woosley2004, Heger2007}. The XRB models of this study reproduce the main feature of the burst light-curve profiles of the clocked bursts and demonstrate that GS~1826$-$24 clocked burster is the candidate exhibiting the extent of nucleosynthesis from the region around CNO cycles up to the SnSbTe cycles found by \citet{Schatz2001}. 
\citet{Koike2004} remarked that the nucleosynthesis of some XRB scenarios may even proceed beyond the SnSbTe cycles and reach $^{126}$Xe if the ignition pressure is greater than $10^{23}$ dyn~cm$^{-2}$.

\begin{figure}[t]
\begin{center}
\includegraphics[width=5.7cm, angle=0]{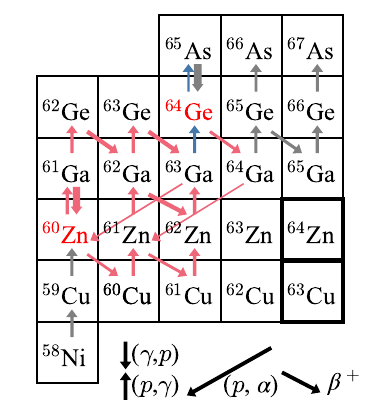}
\caption{\label{fig:Cycles_ZnGa}{The breakout of the ZnGa cycles via the $^{63}$Ga(p,$\gamma$)$^{64}$Ge(p,$\gamma$)$^{65}$As reaction. The arrows depicting the reaction flow are scaled to the logarithmic strength of the respective reaction rates at $T\!=\!1$~GK. The ZnGa cycles are displayed as red arrows. Stable nuclei are represented by thick black squares, and the historical $^{60}$Zn and $^{64}$Ge waiting points are shown in red text. Both $^{63}$Ga(p,$\gamma$)$^{64}$Ge and $^{63}$Ga(p,$\alpha$)$^{60}$Zn reactions compete at $^{63}$Ga. The reaction flow mainly diverges out to $^{64}$Ge via $^{63}$Ga(p,$\gamma$)$^{64}$Ge as the $^{63}$Ga(p,$\gamma$)$^{64}$Ge reaction rate is up to 4 orders of magnitude higher than the $^{63}$Ga(p,$\alpha$)$^{60}$Zn reaction rate at temperature $T\!=\!0.5$-$1.6$~GK (top panel of Fig.~\ref{fig:rp_63Ga_64Ge_60Zn_contri}). Both $^{64}$Ga(p,$\gamma$)$^{65}$Ge and $^{65}$Ga(p,$\gamma$)$^{66}$Ge reactions could be the alternative paths for the reaction flow to reach the weak GeAs cycles. The strengths of these alternative paths are up to a factor of $\sim$13 higher than $^{63}$Ga(p,$\gamma$)$^{64}$Ge at $T\!=\!0.5$-$1.6$~GK (middle and bottom panels of Fig.~\ref{fig:rp_63Ga_64Ge_60Zn_contri}). These alternative paths depend on the strengths of the $\beta$-decay of $^{64}$Ge and of the forward and reverse $^{64}$Ge(p,$\gamma$)$^{65}$As reactions during the onset of XRBs. 
See the detailed study of the weak GeAs cycles performed by \citet{Lam2022a}.}}
\end{center}
\vspace{-5mm}
\end{figure}

\begin{figure}[t]
\includegraphics[width=8.4cm, angle=0]{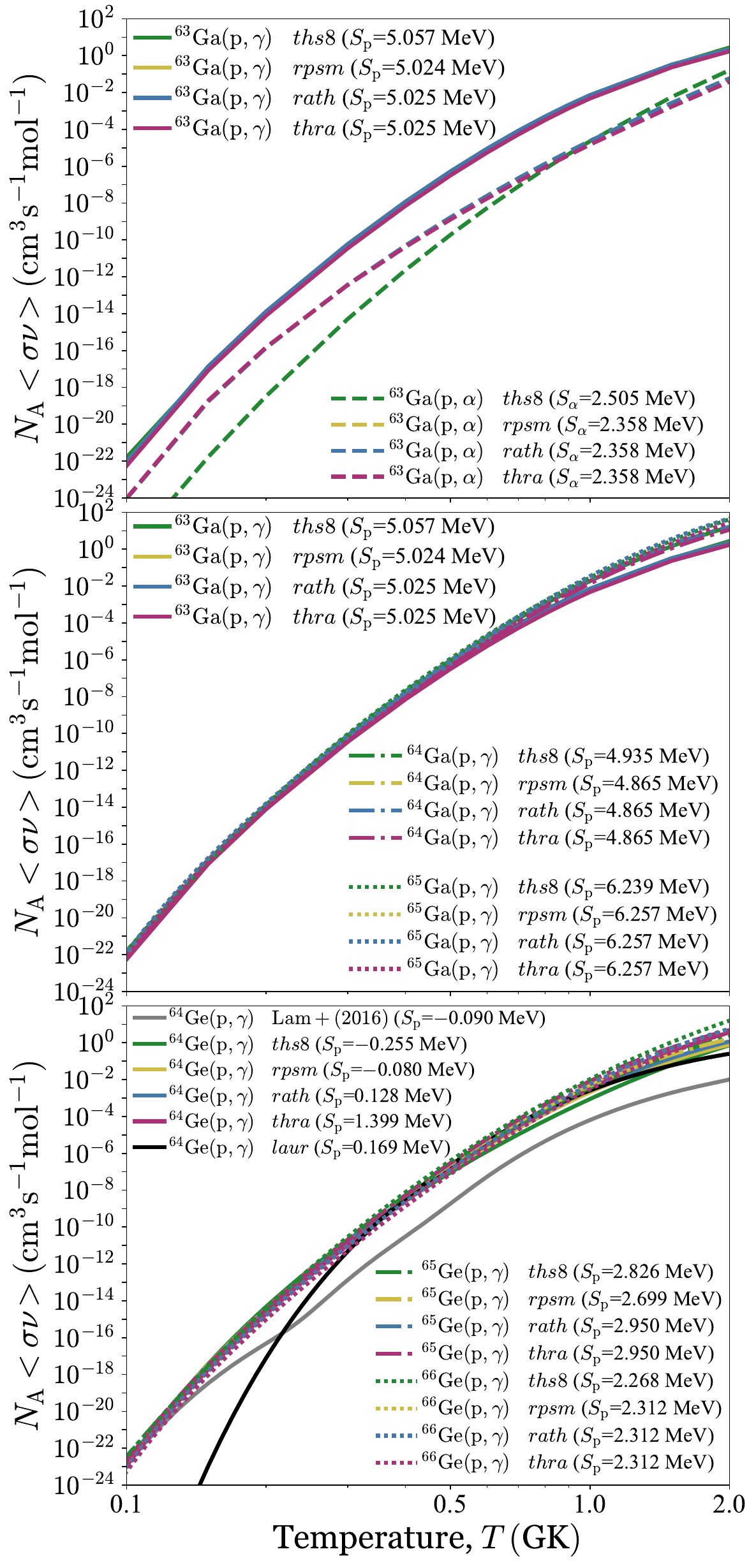}
\caption{\label{fig:rp_63Ga_64Ge_60Zn_contri}{The comparisons of thermonuclear reaction rates connecting the ZnGa and GeAs cycles based on the available reaction rates compiled in JINA REACLIB v2.2 \citep{Cyburt2010}. The temperature range, $0.1\!\leq\!T(\mathrm{GK})\!\leq\!2.0$, \textcolor{black}{is in the region of XRB interest}. Top panel: $^{63}$Ga(p,$\gamma$)$^{64}$Ge and $^{63}$Ga(p,$\alpha$)$^{60}$Zn. Middle panel: $^{63}$Ga(p,$\gamma$)$^{64}$Ge, $^{64}$Ga(p,$\gamma$)$^{65}$Ge, and $^{65}$Ga(p,$\gamma$)$^{66}$Ge. Bottom panel: $^{64}$Ge(p,$\gamma$)$^{65}$As, $^{65}$Ge(p,$\gamma$)$^{66}$As, and $^{66}$Ge(p,$\gamma$)$^{67}$As.}}
\end{figure}

In brief, during the onset of an XRB with temperature $T\!\approx\!0.5$-$2$~GK and density $\rho\!\approx\!10^6$~gcm$^{-3}$, the thermonuclear runaway in the accreted envelope undergoes a series of proton-proton chain and then hot CNO cycles; the reaction flow successively breaks out from the hot CNO cycles via $^{14}$O, $^{15}$O, and $^{18}$Ne through several reaction flows involving $\alpha$p- and $\alpha$-capture processes, i.e., \\
(i) $^{14}$O($\alpha$,p)$^{17}$F(p,$\gamma$)$^{18}$Ne($\alpha$,p)$^{21}$Na(p,$\gamma$)$^{22}$Mg, \\
(ii) $^{14}$O($\alpha$,p)$^{17}$F(p,$\gamma$)$^{18}$Ne($\alpha$,$\gamma$)$^{22}$Mg, \\
(iii) $^{14}$O($\alpha$,p)$^{17}$F(p,$\gamma$)$^{20}$Ne($\alpha$,$\gamma$)$^{24}$Mg, \\ 
(iv)~$^{15}$O($\alpha$,$\gamma$)$^{19}$Ne($\alpha$,p)$^{22}$Na(p,$\gamma$)$^{23}$Mg, \\ 
(v)~$^{12}$C($\alpha$,$\gamma$)$^{16}$O($\alpha$,$\gamma$)$^{18}$Ne($\alpha$,p)$^{21}$Na(p,$\gamma$)$^{22}$Mg, and \\
(vi) $^{12}$C($\alpha$,$\gamma$)$^{16}$O($\alpha$,$\gamma$)$^{18}$Ne($\alpha$,$\gamma$)$^{22}$Mg, \\
and diverges at branching points, e.g., $^{22}$Mg, $^{23}$Mg, and $^{24}$Mg, then reaches the \emph{pf}-shell nuclei after passing through several successive $\alpha$p-, $\alpha$-capture, and rp-processes in the \emph{sd}-shell nuclei \citep{Wormer1994,Wiescher2007}. Both influential $^{22}$Mg($\alpha$,p) and $^{22}$Mg(p,$\gamma$) reactions compete at the $^{22}$Mg branch point, deciding the subsequent reaction-flow paths \citep{Wormer1994,Hu2021}. When the rp-process reaction flow approaches the proton drip line, successive capture of protons by freshly synthesized proton-rich nuclei is inhibited by strong reverse photodisintegrations. Meanwhile, $\beta$ decays constrain the nuclei from further capturing protons as well. After the reaction flow breaks out from the NiCu cycles \citep{Wormer1994,Lam2022b}, it further breaks out from the ZnGa cycles via the $^{63}$Ga(p,$\gamma$)$^{64}$Ge(p,$\gamma$)$^{65}$As path (see Fig.~\ref{fig:Cycles_ZnGa} for the breakout path of the ZnGa cycles and Fig.~\ref{fig:rp_63Ga_64Ge_60Zn_contri} for the respective reaction rates). The reaction flow inevitably passes through the $^{64}$Ge waiting point \citep{Tu2011, Lam2016, Zhou2023}, $^{65}$Ge, and $^{66}$Ge, connecting both ZnGa and weak GeAs cycles \citep{Wormer1994,Lam2022a} and then funneling through the region heavier than the Ge and Se isotopes up to the endpoint of nucleosynthesis of an XRB. Hydrogen is intensively burned in the region, where the reaction flow surges through from the \emph{pf}-shell nuclei until the endpoint, generating the nuclear energy powering up the XRBs, corresponding to the burst light curves. 



The quantitative reproduction of the average observed GS~1826$-$24 clocked burst \textcolor{black}{light-curve profile} is obtained by defining the mixed H/He stellar fuel for the unstable He ignition \citep{Heger2007,Galloway2017}, agreeing with the ``case 1'' semianalytical study of \citet{Fujimoto1981} or the ``prompt mixed'' bursts in regime 3 of \citet{Narayan2003}. The mass fractions of accreted H and He are found to be $\approx\!0.7$ and $\approx\!0.28$, respectively \citep{Heger2007,Galloway2017}. For the case 1 scenario, the \textcolor{black}{accretion} rate ($\sim\!\!0.1\dot{m}_\mathrm{Edd}$, adopting the Eddington-limited accretion rate, $\dot{m}_\mathrm{Edd}\!=\!8.8\times10^4$~gcm$^{-2}$s$^{-1}$) has to be larger than the \textcolor{black}{accretion} rate of the steady state or overcomes the consumption rate of nuclear fuel in the accreted envelope. Such a situation causes the base of the accreted envelope to be driven into the unstable helium-burning region before the accreted hydrogen is depleted in a time $t_\mathrm{CNO}$, developing a mixture of helium burning and proton captures on the synthesized nuclei from helium burning \citep{Fujimoto1981}. Therefore, the recurrence time, $\Delta t_\mathrm{rec}$, of an epoch of GS~1826$-$24 clocked bursts is less than $t_\mathrm{CNO}$ \citep{Lampe2016,Galloway2017}.

The other scenario is ``case 2'' of \citet{Fujimoto1981}, of which the recurrence time may be higher than $t_\mathrm{CNO}$ and the accretion rate is lower than $\sim\!\!0.1\dot{m}_\mathrm{Edd}$. In this case, a lower accretion rate allows the hydrogen-shell burning at the base of the accreted envelope to settle into a steady state, forming the pure helium layer below the hydrogen-burning shell. When the mass of the pure helium layer develops until the ignition threshold, this triggers the helium-shell flash and leads to the ignition in a pure He layer. The pioneer studies by \citet{Johnston2018} and by \citet{Goodwin2019} on the SAX~J1808.4$-$3658 photospheric radius expansion (PRE) burster using the \textsc{Kepler} code show that the (consecutive) PRE bursts are the candidates demonstrating the case 2 scenario. In addition, once a PRE XRB is ignited, the accreted hydrogen for case 2 is depleted much faster than the one for case 1, limiting the rp-process path only up to the region of around nuclear masses, $A\!=\!60$-$64$. Therefore, the GeAs cycles might play an important role in characterizing the PRE burst light curves, burst fluences, and recurrence times.

The first sensitivity study on XRBs based on the postprocessing model was performed by \citet{Parikh2008, Parikh2009}. They employed three sets of hydrodynamic snapshots, i.e., the snapshots of temperature-density profiles, which were generated from the one-zone XRB model with hydrodynamics \citep{Koike2004}, multizone XRB models with hydrodynamics \citep{Fisker2008}, or the one-zone XRB model without hydrodynamics \citep{Schatz2001}. The study investigated the impact of reaction-rate variations on the nucleosynthesis of XRBs. They found that the $^{63}$Ga(p,$\gamma$)$^{64}$Ge and $^{65}$As(p,$\gamma$)$^{66}$Se affect the productions of nearby isotopes, e.g., $^{63}$Cu, $^{64}$Zn, $^{65}$Zn, $^{67}$Ga, $^{66}$Ge, and $^{68}$Ge. Later, according to
the sensitivity studies performed by \citet{Cyburt2016} and \citet{Jacobs2018} on the GS~1826$-$24 clocked burster \citep{Makino1988, Tanaka1989, Ubertini1999}, the $^{63}$Ga(p,$\gamma$)$^{64}$Ge reaction is listed as the fourth key influential (p,$\gamma$) reaction that affects \textcolor{black}{the energy release and the periodic burst light-curve profile of the GS~1826$-$24 clocked burster. And yet only the production of $A\!=\!63$ nuclei is influenced by $^{63}$Ga(p,$\gamma$)$^{64}$Ge.} The advantage of the XRB models used by \citet{Cyburt2016} and \citet{Jacobs2018} is that these models closely, though not exactly, reproduce the averaged-observed periodic light curve of the GS~1826$-$24 clocked burster based on the self-consistent one-dimensional multizone thermohydrodynamics \citep{Heger2007}, which considers the mutual influence between nuclear reactions and thermohydrodynamics. 
\textcolor{black}{Moreover, the treatment of hydrodynamics, such as energy transport through convection and mixing, is also considered in the one-dimensional multizone XRB model. Nevertheless, this essential self-consistent feature is not taken into account in the postprocessing model used by \citet{Parikh2008, Parikh2009} and the one-zone XRB model used by \citet{Schatz2001}, \citet{Schatz2017}, and \citet{Kimura2025}. In addition, the one-zone XRB model is, so far, not capable of reproducing any observational burst properties.} Therefore, a set of hydrodynamic snapshots produced from the one-dimensional multizone XRB model capable of close matching theoretical burst light curves, fluences, peaks, and recurrence times with observations throughout the evolution of the accreted envelope is indeed needed for the postprocessing and one-zone models to acquire the nuclear energy generation, which is consistent with the input hydrodynamic snapshots, and to cross-check the inventory of abundances and the reaction flow during the XRBs \citep{Lam2022a}. The postprocessing and one-zone models without self-consistency are incapable of assessing the burst recurrence times and fluences of a series of XRBs in an epoch that can be done in the one-dimensional multizone XRB models. Such an assessment is very sensitive to the interplay between nuclear reactions, compositional inertia, and thermohydrodynamics. The assessment permits us to study the consequences of this interplay at the duration of several consecutive bursts in detail. 

A recent study of the proton capture on the $^{64}$Ge waiting point was performed by \citet{Zhou2023}, based on the newly measured nuclear masses of $^{63}$Ge, $^{64,65}$As, and $^{66,67}$Se. A signature of reduction in the rp-process at $^{64}$Ge is indicated by them, and the influence of these new masses on the periodic burst light-curve profile of the GS~1826$-$24 clocked burster is surprisingly pronounced. In fact, these new masses mainly affect the proton thresholds of these four (p,$\gamma$) reverse reaction rates: 
$^{63}$Ge(p,$\gamma$)$^{64}$As, 
$^{64}$Ge(p,$\gamma$)$^{65}$As, 
$^{65}$As(p,$\gamma$)$^{66}$Se, and
$^{66}$As(p,$\gamma$)$^{67}$Se. 
However, according to the \citet{Cyburt2016} sensitivity study based on the periodic burst light curve of the GS~1826$-$24 burster, out of these four (p,$\gamma$) reactions, only $^{65}$As(p,$\gamma$)$^{66}$Se is identified as sensitive reaction, provided that it is 2 orders of magnitude different from the recommended reaction. Although the recent sensitivity study by \citet{Schatz2017} based on the one-zone XRB model highlights the potential importance of the $^{66}$Se and $^{65}$As proton thresholds in influencing the burst light curve, the finding is, in fact, only based on a set of burst light curves, and yet the model is not constrained to reproducing any observed burst light curves, fluences, and recurrence times due to the missing essential feature of self-consistency aforementioned. The outcomes of such a sensitivity study could be questionable. In fact, studies based on a set of high-fidelity, well-constrained, and self-consistent models capable of reproducing the observables of the selected bursters are needed to reveal the correlations between variations in thermonuclear reaction rates and observables associated with astrophysical parameters.

Both sensitivity studies performed by \citet{Cyburt2016} and \citet{Jacobs2018} agree that the periodic XRBs of the GS~1826$-$24 burster are not sensitive to $^{63}$Ge(p,$\gamma$)$^{64}$As and $^{66}$As(p,$\gamma$)$^{67}$Se reactions. What we learn from the recent study on the $^{55}$Ni(p,$\gamma$)$^{56}$Cu reaction bypassing the important $^{56}$Ni waiting point \citep{Lam2022b} is that $^{63}$Ge(p,$\gamma$)$^{64}$As is somehow analogous to $^{55}$Ni(p,$\gamma$)$^{56}$Cu, of which the $^{63}$Ge(p,$\gamma$)$^{64}$As reaction is claimed \textcolor{black}{by \citet{Zhou2023} that it induces} a bypass flow from the $^{64}$Ge waiting point. For the case of bypassing the $^{56}$Ni waiting point, the impact of the bypass flow is suppressed by the NiCu cycle, and thus a similar and even stronger suppression on the bypass flow occurs in the GeAs cycle, diminishing the impact of the new $^{63}$Ge and $^{64}$As masses on the periodic XRBs. The influence of the new $^{67}$Se mass on the $^{66}$As(p,$\gamma$)$^{67}$Se forward reaction rate is somehow negligible, as the updated mass is merely $31\!\pm\!70$~keV  different from AME2020 \citep{AME2020}.

Furthermore, the study of \citet{Zhou2023} of the impact of the newly measured masses of $^{63}$Ge, $^{64,65}$As, and $^{66,67}$Se on the periodic XRBs was merely based on the newly deduced proton thresholds of these isotopes without taking into account the respective nuclear structure information for the forward and reverse (p,$\gamma$) reaction rates. In fact, \textcolor{black}{a detailed study of} the impact of the new $^{65}$As and $^{66}$Se proton thresholds on XRBs basically requires the reconstruction of the respective $^{64}$Ge(p,$\gamma$)$^{65}$As and $^{65}$As(p,$\gamma$)$^{66}$Se forward and reverse reaction rates. For instance, the impact of the proton threshold of $^{66}$Se on the $^{65}$As(p,$\gamma$)$^{66}$Se forward and reverse reaction rates and on the periodic XRBs of the GS~1826$-$24 clocked burster had been recently explored in detail by \citet{Lam2022a} prior to the Zhou et al.~work. The proton threshold of $^{66}$Se had been predicted by using the experimental mirror nuclear masses and theoretical mirror displacement energies based on the self-consistent relativistic Hartree-Bogoliubov theory. The theoretical $^{66}$Se proton threshold, $S_\mathrm{p}(^{66}$Se$)\!=\!2.469\!\pm\!0.054$~MeV, agrees well with the experimentally deduced $S_\mathrm{p}(^{66}$Se$)\!=\!2.465\!\pm\!0.074$~MeV. Then, the $^{65}$As(p,$\gamma$)$^{66}$Se forward and reverse reaction rates are reconstructed, and the influence of $S_\mathrm{p}(^{66}$Se$)$ on the $^{65}$As(p,$\gamma$)$^{66}$Se reaction rate and on the periodic burst light-curve profile of the GS~1826$-$24 burster (epoch 1998 June) have been examined in detail. 

As $^{63}$Ga(p,$\gamma$)$^{64}$Ge, $^{64}$Ge(p,$\gamma$)$^{65}$As, and $^{64}$Ge waiting point correlate with the GeAs cycle, which potentially influences the SAX~J1808.4$-$3658 PRE bursts and perhaps the GS~1826$-$24 periodic bursts, we use a new set of $^{63}$Ga(p,$\gamma$)$^{64}$Ge(p,$\gamma$)$^{65}$As reaction rates from \citet{Lu2024} based on the most updated experimental input, e.g., the $^{64}$Ge proton threshold, $S_\mathrm{p}($$^{64}$Ge$)\!=\!5.057\!\pm\!0.004$~MeV (AME2020; \citealt{AME2020}), and the $^{65}$As proton threshold,  $S_\mathrm{p}($$^{65}$As$)\!=\!-0.221\!\pm\!0.042$~MeV \citep{Zhou2023}, and the scarce experimental spectroscopic information in the respective Gamow energies is supplemented with the theoretical information generated from the full \emph{pf}-shell space nuclear shell model calculations with identified uncertainties. As the $^{66}$Se proton threshold, $S_\mathrm{p}($$^{66}$Se$)$, predicted by \citet{Lam2022a} is well matched with the latest experimental finding, $S_\mathrm{p}($$^{66}$Se$)\!=\!2.465\!\pm\!0.074$~MeV \citep{Zhou2023}, we just directly use the latest $^{65}$As(p,$\gamma$)$^{66}$Se reaction rate of \citet{Lam2022a} for this work.

Using a set of high-fidelity and well-constrained XRB models based on the self-consistent one-dimensional multizone thermohydrodynamic code, \textsc{Kepler}, for the GS~1826$-$24 clocked bursts and SAX~J1808.4$-$3658 PRE bursts, we investigate the impact of new $^{63}$Ga(p,$\gamma$)$^{64}$Ge and $^{64}$Ge(p,$\gamma$)$^{65}$As by \citet{Lu2024}, together with the recently updated $^{65}$As(p,$\gamma$)$^{66}$Se \citep{Lam2022a}, $^{55}$Ni(p,$\gamma$)$^{56}$Cu \citep{Valverde2019}, $^{55}$Ni(p,$\gamma$)$^{57}$Cu \citep{Kahl2019}, $^{57}$Cu(p,$\gamma$)$^{58}$Zn \citep{Lam2022b}, and $^{22}$Mg($\alpha$,p)$^{25}$Al \citep{Hu2021} reaction rates on the direct observables of these two bursters, i.e., burst light curves, recurrence times, and fluences. For the present work, the rest of the reaction rates are taken from JINA REACLIB v2.2. The impact of new $^{63}$Ga(p,$\gamma$)$^{64}$Ge and $^{64}$Ge(p,$\gamma$)$^{65}$As reaction rates on the XRBs of the periodic and PRE bursts is presented in the following section, and the conclusion of this work is given in Section~\ref{sec:summary}.

\newpage
\section{Influence of new $^{63}$G\lowercase{a}(\lowercase{p},$\gamma$)$^{64}$G\lowercase{e} and $^{64}$G\lowercase{e}(\lowercase{p},$\gamma$)$^{65}$A\lowercase{s} reaction rates on the XRBs of the periodic and PRE bursts}
\label{sec:astrophysical_implications}


The essential feature of the periodic bursts of the GS~1826$-$24 burster is the unusually regular burst light curves and almost constant recurrence times, causing it to be dubbed the ``clocked'' burster \citep{Ubertini1999, Galloway2004}. This regular bursting behavior implies that the accretion rate mildly varies on short timescales, within the range of 5-13~\%~$\dot{m}_\mathrm{Edd}$ \citep{Galloway2008, Galloway2017}. The periodic bursts of the GS~1826$-$24 burster are the most investigated because of the nearly constant accretion rate and consistent burst light-curve profile.

In 2002 October, the SAX~J1808.4$-$3658 burster ignited a series of XRBs having irregular burst patterns and nonconstant recurrence times compared to the periodic bursts of GS~1826$-$24 burster. These bursts are categorized as PRE bursts \citep{Galloway2004, Galloway2008}. The PRE bursts of this epoch have a typical rapid rise in X-ray luminosity ($\lessapprox\!1$~s) followed by a wide plateau ($\approx\!10$~s) and a short tail ($\lessapprox\!30$~s; \citealt{Johnston2018}), of which the local X-ray luminosity may reach the Eddington limit, and the enormous expansion of photospheric layers is lifted off the neutron star surface by radiation pressure, featuring an approximately constant total X-ray flux when the blackbody temperature decreases while the inferred blackbody radius increases \citep{Strohmayer2006}.

These two XRB scenarios of different burst features could possibly reveal the influence of the newly determined $^{63}$Ga(p,$\gamma$)$^{64}$Ge and $^{64}$Ge(p,$\gamma$)$^{65}$As reaction rates. The study of these reaction rates is arranged in two groups, i.e., \\
(i) $^{63}$Ga(p,$\gamma$)$^{64}$Ge and $^{64}$Ge(p,$\gamma$)$^{65}$As from \citet{Lu2024} and $^{22}$Mg($\alpha$,p)$^{25}$Al from JINA REACLIB v2.2 \citep{Cyburt2010} and \\
(ii) $^{63}$Ga(p,$\gamma$)$^{64}$Ge and $^{64}$Ge(p,$\gamma$)$^{65}$As from \citet{Lu2024} and $^{22}$Mg($\alpha$,p)$^{25}$Al from \citet{Hu2021}. \\ 
Updated reaction rates, i.e., $^{55}$Ni(p,$\gamma$)$^{56}$Cu \citep{Valverde2019}, $^{56}$Ni(p,$\gamma$)$^{57}$Cu \citep{Kahl2019}, $^{65}$As(p,$\gamma$)$^{66}$Se \citep{Lam2022a}, and $^{57}$Cu(p,$\gamma$)$^{58}$Zn \citep{Lam2022b}, are also incorporated in both group (i) and group (ii), and the rest of the reaction rates are from JINA REACLIB v2.2. 
Hereinafter, the models implementing group (i) and group (ii) reaction rates are named \emph{Present}$^\dag$ and \emph{Present}$^\ddag$, respectively. Unless otherwise specified, the model using only JINA REACLIB v2.2 is labeled as ``baseline''.

For the group (ii) reaction rates, we adopt the $^{22}$Mg($\alpha$,p)$^{25}$Al reaction rate by \citet{Hu2021}, in which the nuclear resonance properties were experimentally studied at the energies corresponding to the XRB Gamow window. Recently, two direct measurements have also been performed for this reaction by \citet{Randhawa2020} and \citet{Jayatissa2023}, but their evaluated reaction rates disagree with each other by a factor of around $4$ for an unknown reason. Moreover, the lowest measured energy was $3.2$~MeV in both measurements, corresponding to $3.1$~GK. This temperature (e.g., \citealt{Randhawa2020}) is still much higher than the characteristic peak temperature ($2$~GK) or the temperature range of either periodic or PRE XRBs, i.e., $0.7$-$1.8$~GK. They relied on the extrapolation by a statistical model calculation to reach the lower temperature, but the validity of the statistical model at the low energy, where the level density is also lower, can be questioned. Note that the reaction rate we adopt was fairly close to the value from one of the direct measurements by \citet{Randhawa2020} for the relevant temperature range. Future direct measurements for the $^{22}$Mg($\alpha$,p)$^{25}$Al reaction in the Gamow window corresponding to the XRB temperature regime are certainly useful in preventing introduction of the unknown uncertainties imposed from the model-dependent extrapolation.


\begin{figure*}[t]
\begin{center}
\includegraphics[width=0.8\textwidth, angle=0]{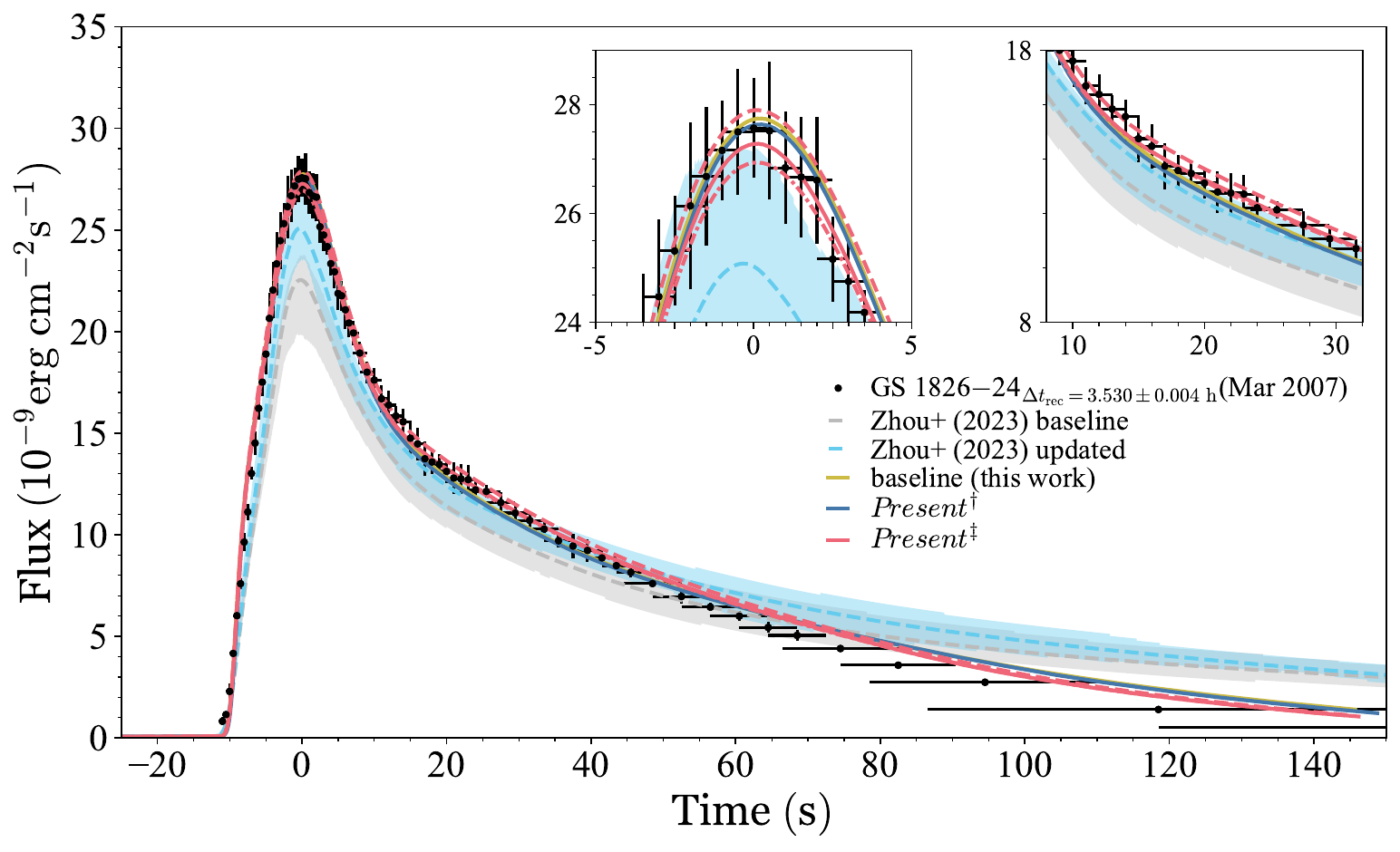}
\caption{\label{fig:periodic_flux}The observed and modeled burst light-curve profiles of the GS~1826$-$24 clocked burster (epoch Mar 2007). The observed burst light-curve profile (black dots) is plotted against the respective modeled counterparts. Modeled light-curve profiles from \citet{Zhou2023}: baseline (dashed gray line) and updated (dashed cyan line). The light-curve profiles of the upper and lower limit of group (i) reaction rates are indiscernible from the respective centroid light curve (\emph{Present}$^\dag$; blue line). The burst peaks of the upper limit (dashed red line) and lower limit (dash-dotted red line) of group (ii) reaction rates enclose the burst peak of the respective centroid light-curve profile (\emph{Present}$^\ddag$; solid red line). The burst peak of \emph{Present}$^\dag$ is merely \textcolor{black}{$\approx\!0.4$~\%} different from the burst peak of baseline of this work (yellow line).}  
\end{center}
\end{figure*}

\begin{figure}[t]
\begin{center}
\includegraphics[width=0.8\columnwidth, angle=0]{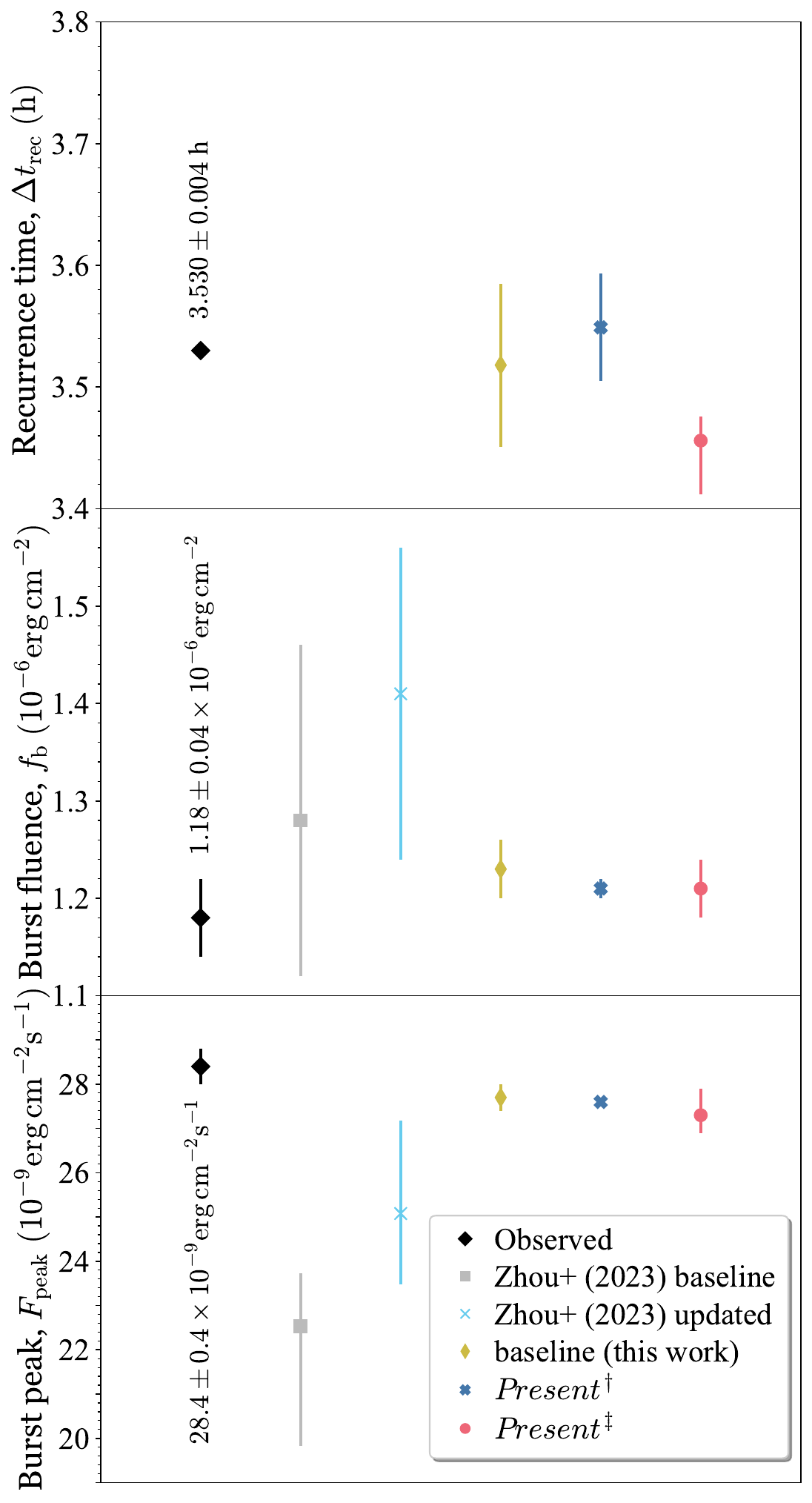}
\caption{\label{fig:periodic_observables}The observed and modeled burst recurrence times, fluences, and peaks of the GS~1826$-$24 clocked burster (epoch 2007 March). The observed burst properties (black diamonds) are plotted against the respective modeled counterparts, i.e., the baseline of this work (yellow diamonds), \emph{Present}$^\dag$ (blue crosses), and \emph{Present}$^\ddag$ (red dots). The burst fluences and peaks from \citet{Zhou2023}, baseline (gray squares) and updated (cyan crosses), are calculated in this work.}
\end{center}
\end{figure}

\subsection{GS~1826$-$24 clocked burster}
\label{sec:periodic}

We construct the GS~1826$-$24 clocked burst model matching with the properties of the periodic XRBs of epoch 2007 March using the \textsc{Kepler} code \citep{Weaver1978, Woosley2004, Heger2007}. The astrophysical configuration of the GS~1826$-$24 model from \citet{Jacobs2018} is adopted by assigning the accreted $^1$H, $^4$He, and CNO metallicity fractions as $0.76$, $0.2275$, and $0.0125$, respectively, while the accretion rate is adjusted to a factor of $0.126$ of the adopted Eddington-limited accretion rate, $\dot{m}_\mathrm{Edd}\!=\!8.8\!\times\!10^4$g$\cdot$cm$^{-2}$s$^{-1}$. The mass and radius of the neutron star are defined as 1.7~$M_\odot$ and 10.84~km, respectively. The base luminosity is 0.025~MeV nucleon$^{-1}$. The redefined astrophysical configuration permits us to construct the initial model capable of reconstructing the averaged-observed bolometric light-curve profile and recurrence time of epoch 2007 March, determined from the time-resolved spectral analysis by \citet{Galloway2017}. The observational data were recorded by the Proportional Counter Array \citep{Jahoda1996} and All Sky Monitor \citep{Levine1996} of the Rossi X-ray Timing Explorer (RXTE; \citealt{Galloway2004,Galloway2008}) and recently compiled into the Multi-Instrument Burst Archive \citep{MINBAR}. These observed periodic burst light curves were stacked up and averaged by Galloway et al.~to produce the burst light-curve profile to ease the comparison among models and observation \citep{Galloway2017}.

We implement the simulation procedure used by \citet{Hu2021} and \citet{Lam2022a,Lam2022b} in which a series of 40 consecutive XRBs for the models based on either the JINA REACLIB v2.2 or group (i) or group (ii) aforementioned reaction rates are calculated and named as the baseline (the initial model), \emph{Present}$^\dag$, and \emph{Present}$^\ddag$, respectively.

The first 10 bursts produced from every model are not taken into account because these 10 bursts are in transition from a chemically fresh envelope to an enriched envelope with stable burning. Moreover, the burned-in burst ashes from the 10th burst onward are somehow more consistently deposited and recycled in the following burst heatings and stabilize the succeeding bursts. Hence, we stack up and average the last 30 bursts to produce a modeled burst light-curve profile. Our averaging procedure is similar to that applied by \citet{Galloway2017} to yield the averaged-observed light-curve profile of epoch 2007 March. The same averaging procedure was implemented in the recent works of \citet{Hu2021} and \citet{Lam2022a,Lam2022b}.

Each modeled burst luminosity, $L_\mathrm{x}$, is transformed to the respective modeled flux, $F_\mathrm{x}$, via the relation $F_\mathrm{x}\!=\!L_\mathrm{x}/\left[ 4\pi d^2\xi_b (1\!+\!z)^2 \right]$, where $d$ is the distance and $\xi_b$ considers the possible deviation of the observed flux from an isotropic burster luminosity \citep{Fujimoto1988,He2016}. The anisotropy factors of the burst and persistent emissions degenerate with distance, and thus $d$ and $\xi_b$ can be combined to form the modified distance, $d\sqrt{\xi_b}$. Both $d\sqrt{\xi_b}$ and redshift ($z$) factors are highly sensitive in regulating the burst light-curve profile when transforming into an observer's frame.

Following the procedure of \citet{Hu2021} and \citet{Lam2022a,Lam2022b}, we treat the baseline, \emph{Present}$^\dag$, and \emph{Present}$^\ddag$ models on an equal footing in obtaining the best match with the directly observed properties, i.e., burst light-curve profile, fluence, and recurrence time. We impartially fit the modeled burst light-curve profile of each model to the observed counterpart of the entire burst time span. This approach allows us to prevent artifactually reshaping the modeled burst light curve and shifting the modeled burst peak or even to avoid imposing subjective selection on the $d\sqrt{\xi_b}$ and $(1\!+\!z)$ factors, which are very sensitive in characterizing the modeled burst light-curve profile. We can avoid imposing an unknown uncertainty on the modeled burst light curve and further inference from the properties of modeled XRBs. Furthermore, the impact of the reaction rates of groups (i) and (ii) on the burst fluence and recurrence time can also be objectively revealed.

For each model of this work, we obtain a set of $d\sqrt{\xi_b}$ and $(1\!+\!z)$ factors by matching the respective modeled burst light-curve profile with the observed light-curve profile of epoch 2007 March. These $d\sqrt{\xi_b}$ and $(1\!+\!z)$ factors are selected by the reproduction of the observed burst light-curve profile with the lowest range of average deviation between the modeled and observed light curve, which is limited to 
$0.55\!\times\!10^{-9}$erg$\cdot$cm$^{-2}$s$^{-1}$ for baseline, 
$0.52\!\times\!10^{-9}$erg$\cdot$cm$^{-2}$s$^{-1}$ for \emph{Present}$^\dag$, and
$0.45\!\times\!10^{-9}$erg$\cdot$cm$^{-2}$s$^{-1}$ for \emph{Present}$^\ddag$. 
Meanwhile, these factors are also constrained by the reproduction of the observed recurrence time ($\Delta t_\mathrm{rec}\!=\!3.530\!\pm\!0.004$~h) with allowing the uncertainty up to $\approx\!2$~\%.

The modeled and observed burst light-curve profiles are compared with one another in Fig.~\ref{fig:periodic_flux}. The best-fit average $d\sqrt{\xi_b}$ and $(1\!+\!z)$ factors of the baseline, \emph{Present}$^\dag$, and \emph{Present}$^\ddag$ modeled light curves constrained to the averaged-observed light curve and recurrence time of epoch 2007 March are 
$7.90\!\pm\!{0.16}$~kpc and $1.04\!\pm\!{0.02}$,
$7.92\!\pm\!{0.10}$~kpc and $1.05\!\pm\!{0.02}$, and 
$7.88\!\pm\!{0.04}$~kpc and $1.04\!\pm\!{0.01}$, respectively. With the best-fit $(1\!+\!z)$ factors, we obtain the respective modeled recurrence times of 
$\Delta t_\mathrm{rec}\!=\!3.518\pm0.067$~hr, 
$\Delta t_\mathrm{rec}\!=\!3.549\pm0.044$~hr, and 
$\Delta t_\mathrm{rec}\!=\!3.456\pm0.020$~hr, 
for the baseline, \emph{Present}$^\dag$, and \emph{Present}$^\ddag$, respectively. 
The determination of the mean values and uncertainties of $d\sqrt{\xi_b}$, $(1\!+\!z)$, and the modeled recurrence times of \emph{Present}$^\dag$ (or \emph{Present}$^\ddag$) have taken into account the centroid, upper, and lower limits of the reaction rates of group (i) or group (ii).

The modeled and observed recurrence times are displayed in the top panel of Fig.~\ref{fig:periodic_observables}. The reaction rates of group (i) used in \emph{Present}$^\dag$ delay the modeled recurrence time about $0.031$~hr ($\approx\!0.9$~\%) compared to the recurrence time of the baseline, whereas for \emph{Present}$^\ddag$ using the reaction rates of group (ii), the inclusion of the experimentally determined $^{22}$Mg($\alpha$,p)$^{25}$Al reaction rate that is more than 1 order of magnitude lower than the recommended NON-SMOKER rate, permits more reaction flow funneling to the $^{22}$Mg(p,$\gamma$)$^{23}$Al(p,$\gamma$)$^{24}$Si($\alpha$,p) branch, speeds up the depletion of accreted hydrogen, and prepares the base of the accreted envelope for the onset of succeeding periodic XRBs. This process shortens about $0.062$~hr ($\approx\!2$~\%) of the recurrence time compared to the one of baseline. 
Overall, the modeled-averaged burst recurrence times of baseline and \emph{Present}$^\dag$ are in good agreement with the observed-averaged recurrence time, i.e., $\Delta t_\mathrm{rec}\!=\!3.530\!\pm\!0.004$~hr, whereas the modeled-averaged burst recurrence time of \emph{Present}$^\ddag$ is $\approx\!2$~\% offset from the observed one, suggesting that a 1\%-2\% increment can be applied for the accretion rate of \emph{Present}$^\ddag$. 

The burst peak from \emph{Present}$^\dag$ is merely \textcolor{black}{$0.01\!\times\!\mathrm{10}^{-9}$~erg$\cdot$cm$^{-2}$s$^{-1}$} different from the burst peak of the baseline model of this work (blue and yellow lines in Fig.~\ref{fig:periodic_flux}), and the burst tails of both the baseline and \emph{Present}$^\dag$ are not discernible, proposing that the \textcolor{black}{reaction rates of group (i) used in} \emph{Present}$^\dag$ mildly affect the periodic burst light-curve profile. The baseline and updated models of \citet{Zhou2023} produce a set of burst peaks that are up to \textcolor{black}{$6.32\!\times\!\mathrm{10}^{-9}$}~erg$\cdot$cm$^{-2}$s$^{-1}$ \textcolor{black}{($\approx\!22$~\%)} away from the observed burst peak, and the difference between each other is \textcolor{black}{$2.54^{+4.8}_{-0.4}\!\times\!\mathrm{10}^{-9}$}~erg$\cdot$cm$^{-2}$s$^{-1}$ (gray and cyan lines in Fig.~\ref{fig:periodic_flux}, and gray squares and cyan crosses in the bottom panel of Fig.~\ref{fig:periodic_observables}). The present finding is consistent with the sensitivity study of \citet{Cyburt2016} -- $^{64}$Ge(p,$\gamma$)$^{65}$As is not sensitive to the periodic burst light curve--and contradicts with the outcome of \citet{Zhou2023}. This contradictory outcome highlights the importance of constructing the initial baseline and comparison models \textcolor{black}{on an} equal footing and of impartially fitting the modeled burst light curves to the entire burst time span of the aforementioned averaged-observed light curve, which was not considered in the work of \citet{Zhou2023} that specifically selected observational data close to the burst peak at $t\!=\!-10$ to $40$~s as \textcolor{black}{previously done} \citep{Meisel2018}. 
Moreover, note that the importance of constraining the GS~1826$-$24 model on reproducing the burst peak, fluence, and recurrence time is crucial.


For the scenario of reaction rates of group (ii) used in \emph{Present}$^\ddag$, the upper limit of both $^{63}$Ga(p,$\gamma$)$^{64}$Ge and $^{64}$Ge(p,$\gamma$)$^{65}$As coupled with the recent experimentally determined $^{22}$Mg($\alpha$,p)$^{25}$Al reaction rate allows more and early hydrogen burning in the accreted envelope. This situation enhances the burst peak by $0.6\!\times\!\mathrm{10}^{-9}$~erg$\cdot$cm$^{-2}$s$^{-1}$ (dashed red line in Fig.~\ref{fig:periodic_flux} and red dot in bottom panel of Fig.~\ref{fig:periodic_observables}) 
and shortens the recurrence time up to $0.044$~hr (red dot in top panel of Fig~\ref{fig:periodic_observables}). 
For the scenario of group (ii) with the lower-limit $^{63}$Ga(p,$\gamma$)$^{64}$Ge and $^{64}$Ge(p,$\gamma$)$^{65}$As reaction rates, on the one hand, implementing the experimentally determined $^{22}$Mg($\alpha$,p)$^{25}$Al reaction rate opens a faster path of $^{22}$Mg(p,$\gamma$)$^{23}$Al(p,$\gamma$)$^{24}$Si($\alpha$,p) branch than the $^{22}$Mg($\alpha$,p)$^{25}$Al path, but on the other hand, the reaction flow is restricted from passing through the GeAs cycles and surging through the region heavier than Ge and Se isotopes of where the intensive hydrogen burning occurs. The depletion of accreted hydrogen is delayed. Thus, the burst peak is reduced by $0.4\!\times\!\mathrm{10}^{-9}$~erg$\cdot$cm$^{-2}$s$^{-1}$ (dashed-dotted red line in Fig.~\ref{fig:periodic_flux} and red dot in bottom panel of Fig.~\ref{fig:periodic_observables}) 
and the recurrence time is extended by $0.02$~hr (red dot in top panel of Fig~\ref{fig:periodic_observables}).

In the present work, baseline, \emph{Present}$^\dag$, and \emph{Present}$^\ddag$ closely though not exactly, reproduce the decrease of the observed burst tail from $t\!=\!0$-$150$~s. This permits us to obtain a set of modeled burst fluences, which is an integration of flux over time, for baseline, \emph{Present}$^\dag$, and \emph{Present}$^\ddag$, 
$1.23\!\pm0.03\!\times\!10^{-6}$~erg$\cdot$cm$^{-2}$, 
$1.21\!\pm0.01\!\times\!10^{-6}$~erg$\cdot$cm$^{-2}$, and 
$1.21\!\pm0.03\!\times\!10^{-6}$~erg$\cdot$cm$^{-2}$, respectively, 
in good agreement with the observed fluence, $f_b\!=\!1.18\!\pm0.04\!\times\!10^{-6}$~erg$\cdot$cm$^{-2}$ \citep{Galloway2017}. The updated $^{63}$Ga(p,$\gamma$)$^{64}$Ge, $^{64}$Ge(p,$\gamma$)$^{65}$As, and $^{22}$Mg($\alpha$,p)$^{25}$Al improve the \emph{Present}$^\dag$ and \emph{Present}$^\ddag$ models in reproducing a set of burst fluences of about $2$~\% closer to the observed burst fluence than baseline.

The burst fluences of the baseline and updated models of \citet{Zhou2023} are 
$1.28^{+0.18}_{-0.16}\!\times\!10^{-6}$~erg$\cdot$cm$^{-2}$ and 
$1.41^{+0.15}_{-0.17}\!\times\!10^{-6}$~erg$\cdot$cm$^{-2}$,  
respectively. These two fluences are up to around $20$~\% deviated from the observation (gray dot and cyan cross in the middle panel of Fig.~\ref{fig:periodic_observables}). 
After the burst peak at $t\!=\!0$~s, both modeled burst tails clearly deviate from the observed burst tail (gray and cyan lines in Fig.~\ref{fig:periodic_flux}). 
The pronounced mismatch between the observed burst peak and tail and the ones produced from the baseline and updated models of \citet{Zhou2023} is further manifested in the reproduction of the observed burst fluence due to the quality of the burst light-curve fitting, which did not consider the whole burst time span of the average observed light curve. The mismatch of burst peak and tail also renders the influence of the newly measured proton threshold of $^{65}$As on GS~1826$-$24 presented by \citet{Zhou2023} somehow overestimated. Note that the range of $(1\!+\!z)$ factors found by \citet{Zhou2023} was, however, based on the modeled burst light-curve profile mismatched with the observed one. In fact, the impact of this mismatch also affects the inference on the neutron star compactness, which sensitively depends on the redshift, $z$, deduced from the burst fit \citep{Lampe2016,Johnston2020}, 
$(1\!+\!z)\!=\!1/\sqrt{1-2GM_\mathrm{GR}/(c^2 R_\mathrm{GR})}$, 
where $G$ is the gravitational constant, $c$ is the speed of light, and $M_\mathrm{GR}$ and $R_\mathrm{GR}$ are the general relativistic neutron star mass and radius, respectively.

Furthermore, as pointed out by \citet{Hu2021} and \citet{Lam2022a,Lam2022b}, the models used by \citet{Zhou2023} yield a set of modeled burst tails that are somehow overexpanded and do not recede accordingly with the observed counterpart. This issue indicates that the modeled hydrogen burning does not recede accordingly with observation and affects the calculated burst ash composition; see the gray and cyan lines in Fig.~\ref{fig:periodic_flux}, Figure~4 in \citet{Randhawa2020}, or Figure~3 in \citet{Hu2021}. 



\begin{figure*}[t!]
\begin{center}
\includegraphics[width=0.32\textwidth, angle=0]{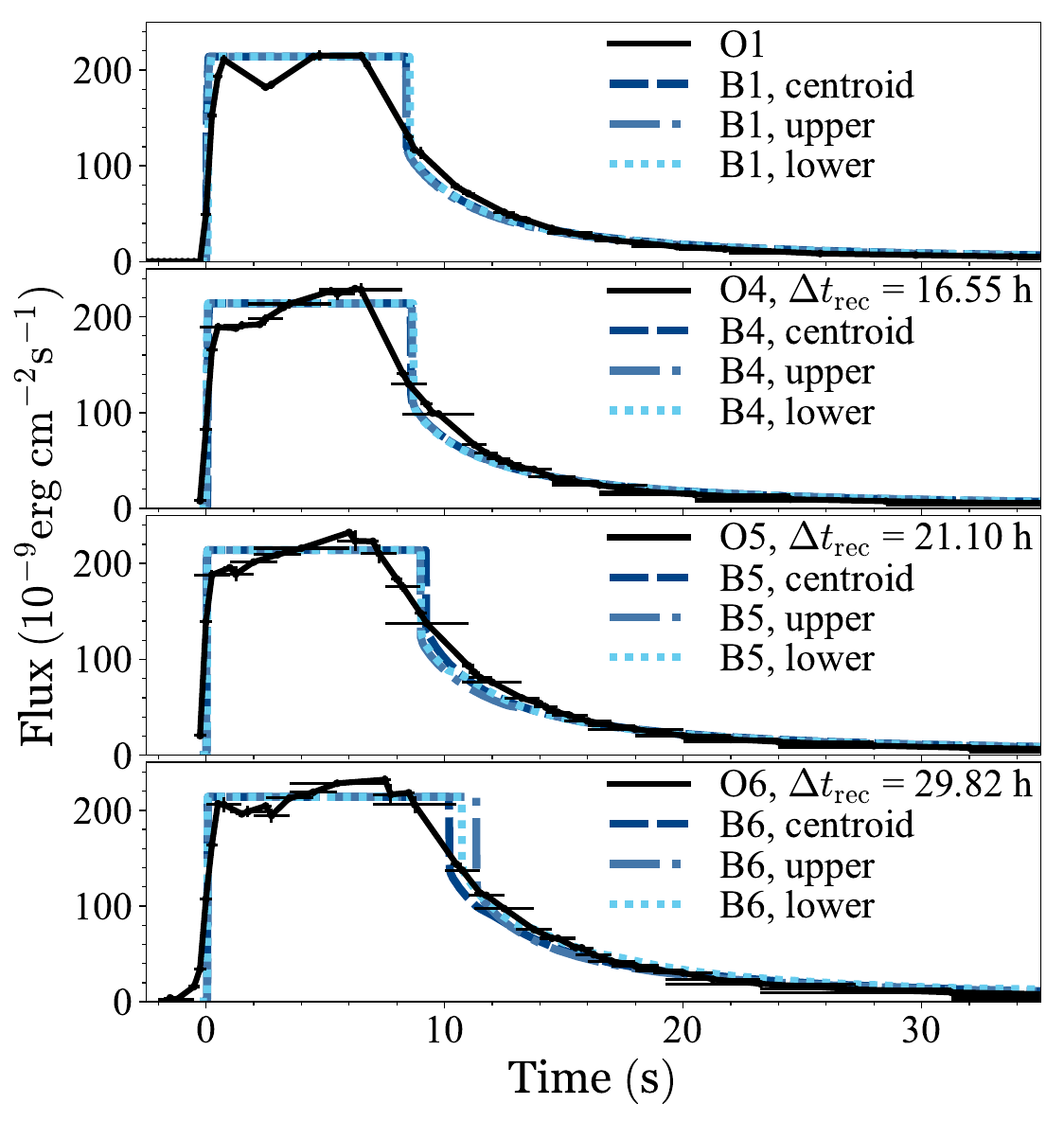}
\includegraphics[width=0.32\textwidth, angle=0]{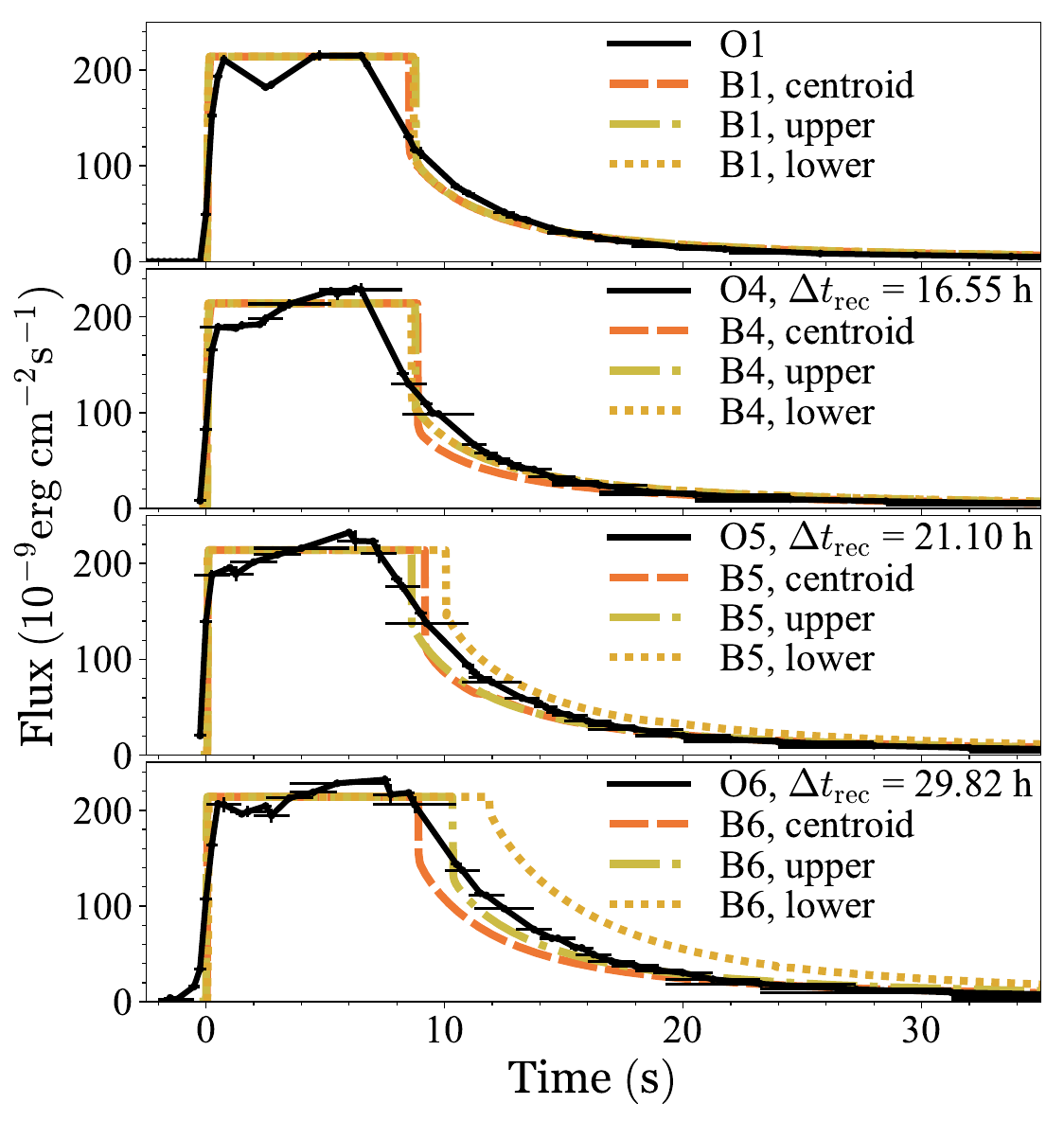}
\includegraphics[width=0.32\textwidth, angle=0]{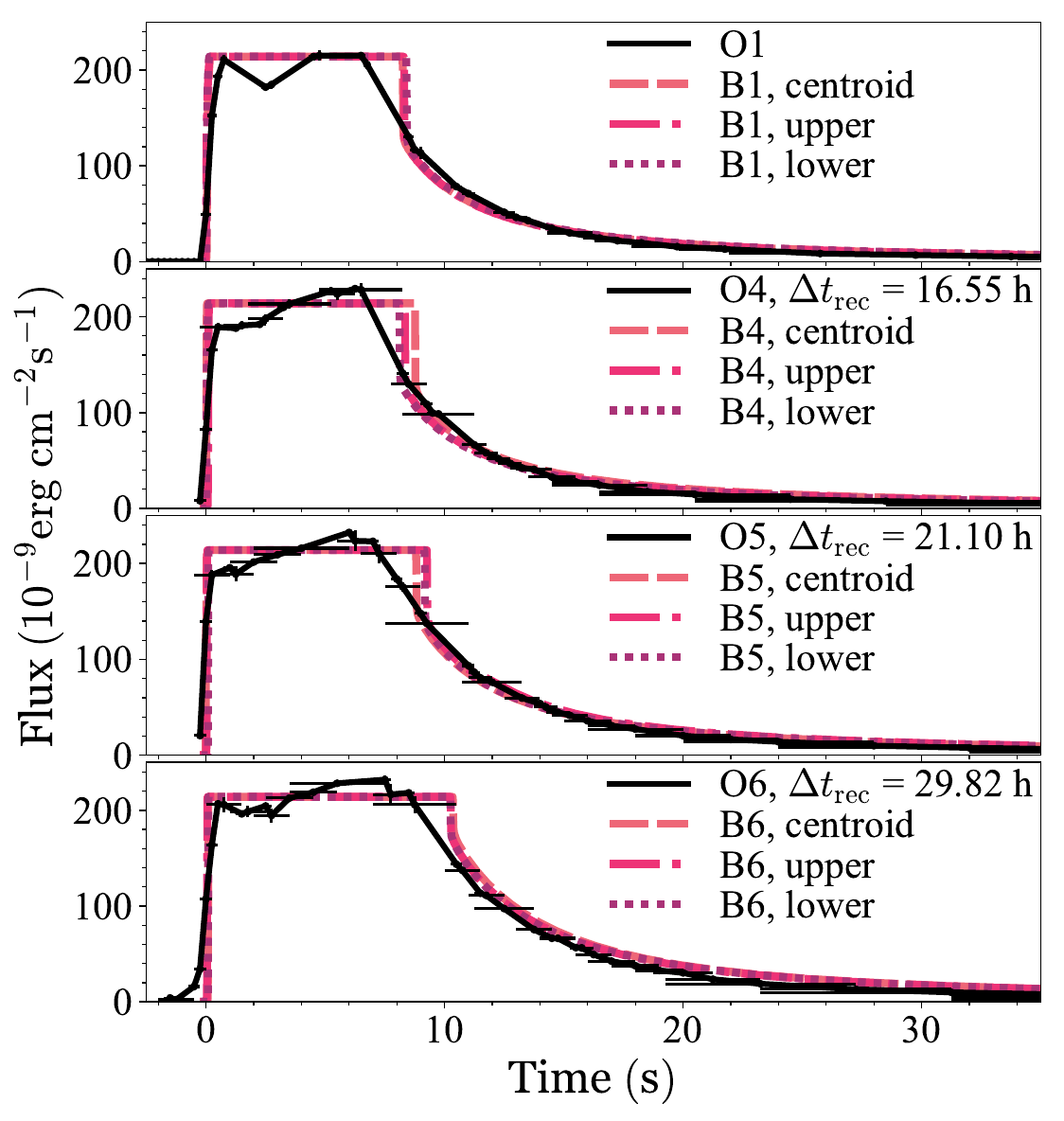}
\caption{\label{fig:PRE_flux}{The observed and modeled PRE burst light curves of SAX J1808.4$-$3658 (epoch 2002 October). Each observed burst light curve of the epoch (black lines) is plotted against the respective modeled counterparts. The modeled fluxes are transformed from the modeled luminosities using the factor $4\pi d^2\xi_b\!\approx\!0.9\!\times\!10^{45}$~cm$^2$ as previously implemented~\citep{Hu2021}. 
Left column: modeled light curves are constructed from the J01 PRE model using the centroid, upper, and lower limits of the reaction rates of group (i). Middle column: modeled light curves yielded from the J01 PRE model using the centroid, upper, and lower limits of the reaction rates of group (ii). Right column: modeled light curves generated from the restored PRE burst model \citep{Hu2021} using the centroid, upper, and lower limits of the reaction rates of group (ii). See text.}}
\end{center}
\end{figure*}

\subsection{SAX~J1808.4$-$3658 PRE burster}
\label{sec:PRE}

The SAX~J1808.4$-$3658 PRE burst model that we use for this work was first constructed by \citet{Johnston2018} and was studied by \citet{Goodwin2019}. The pioneer PRE burst model is based on the \textsc{Kepler} code using JINA REACLIB v2.2 \citep{Cyburt2010}. Implementing the new experimentally deduced $^{22}$Mg($\alpha$,p)$^{25}$Al reaction rate, \citet{Hu2021} studied the influence of the new $^{22}$Mg($\alpha$,p)$^{25}$Al reaction rate on PRE bursts and found a strong correlation between the He abundance in the base of the accreted envelope of the host neutron star of the SAX~J1808.4$-$3658 burster and the dominance of the $^{22}$Mg($\alpha$,p)$^{25}$Al branch. The impact of the new $^{22}$Mg($\alpha$,p)$^{25}$Al reaction rate requires them to further constrain the He abundance in the base to be $X(\mathrm{He})\!=\!0.567\!\pm\!0.003$. This constraint not only restores but also improves the PRE burst model matching the observed burst fluences and recurrence times of epoch 2002 October, which was recorded by the Proportional Counter Array \citep{Jahoda1996} and the All Sky Monitor \citep{Levine1996} of the RXTE \citep{Galloway2004,Galloway2008,MINBAR}. 

The PRE burst models permit us to concurrently study the impact of key reactions on the burst light curves, recurrence times, and fluences. Comparing the PRE bursts to periodic bursts, an epoch of PRE bursts could be more sensitive to the updated reaction rates in important cycles and/or at decisive branch points due to the nonconstant recurrence times and the high intercorrelation between nuclear energy generation and thermohydrodynamics. 

In this work, we employ the \citet{Johnston2018} PRE burst model (hereinafter J01 model) and the restored PRE burst model \citep{Hu2021} to study the impact of two groups of updated reaction rates. 
The former model takes into account the reaction rates of either group (i) or (ii), whereas the latter model is used \textcolor{black}{for studying the reaction rates of group (ii)}. 

\renewcommand{\arraystretch}{0.8}
\begin{table}
\scriptsize
\caption{\label{tab:PRE_fluence}Burst properties yielded from the J01 \citep{Johnston2018} and restored \citep{Hu2021} PRE burst models using the centroid, upper, and lower limits of \textcolor{black}{the reaction rates of group (i) or (ii)}.}
\begin{tabular*}{\linewidth}{@{\hspace{2mm}\extracolsep{\fill}}cclccc@{\hspace{2mm}}}
\toprule
\midrule[0.25pt]
 Burst   & PRE      & Reaction   & $t$  & $\Delta t_\mathrm{rec}$  & Fluence, $f_b$  \\
         & model    & group      & (hr)  & (hr)                    & ($10^{-6}$~erg$\cdot$cm$^{-2}$)  \\
\hline
{\bf O1} &          &               & $0.0$     & --       & $2.620\pm0.021$ \\
{\bf B1} & J01      & (i) centroid  & $0.0$     & --       & $2.529$         \\
         & J01      & (i) upper     & $0.0$     & --       & $2.495$         \\
         & J01      & (i) lower     & $0.0$     & --       & $2.525$         \\
         & J01      & (ii) centroid & $0.0$     & --       & $2.517$         \\
         & J01      & (ii) upper    & $0.0$     & --       & $2.561$         \\
         & J01      & (ii) lower    & $0.0$     & --       & $2.549$         \\
         & Restored & (ii) centroid & $0.0$     & --       & $2.548$         \\
         & Restored & (ii) upper    & $0.0$     & --       & $2.545$         \\
         & Restored & (ii) lower    & $0.0$     & --       & $2.579$         \\
\midrule[0.20pt]
     B2  & J01      & (i) centroid  & $16.2$    & $16.2$   & $2.647$         \\
         & J01      & (i) upper     & $17.1$    & $17.1$   & $2.646$         \\
         & J01      & (i) lower     & $16.2$    & $16.2$   & $2.583$         \\
         & J01      & (ii) centroid & $14.1$    & $14.1$   & $2.454$         \\
         & J01      & (ii) upper    & $13.7$    & $13.7$   & $2.481$         \\
         & J01      & (ii) lower    & $13.8$    & $13.8$   & $2.546$         \\
         & Restored & (ii) centroid & $17.2$    & $17.2$   & $2.561$         \\
         & Restored & (ii) upper    & $16.4$    & $16.4$   & $2.541$         \\
         & Restored & (ii) lower    & $16.1$    & $16.1$   & $2.551$         \\
\midrule[0.20pt]
     B3  & J01      & (i) centroid  & $32.2$    & $16.0$   & $2.574$         \\
         & J01      & (i) upper     & $32.9$    & $15.8$   & $2.512$         \\
         & J01      & (i) lower     & $32.3$    & $16.1$   & $2.582$         \\
         & J01      & (ii) centroid & $30.6$    & $16.5$   & $2.588$         \\
         & J01      & (ii) upper    & $27.1$    & $13.4$   & $2.483$         \\
         & J01      & (ii) lower    & $27.5$    & $13.7$   & $2.479$         \\
         & Restored & (ii) centroid &  --       &  --      &  --             \\
         & Restored & (ii) upper    &  --       &  --      &  --             \\
         & Restored & (ii) lower    &  --       &  --      &  --             \\
\midrule[0.20pt]
{\bf O4} &          &               & $45.4$    & $45.4^d$ & $2.649\pm0.018$ \\
{\bf B4} & J01      & (i) centroid  & $48.5$    & $48.5^d$ & $2.534$         \\
         & J01      & (i) upper     & $49.5$    & $49.5^d$ & $2.560$         \\
         & J01      & (i) lower     & $48.7$    & $48.7^d$ & $2.561$         \\
         & J01      & (ii) centroid & $43.9$    & $43.9^d$ & $2.436$         \\
         & J01      & (ii) upper    & $44.8$    & $44.8^d$ & $2.567$         \\
         & J01      & (ii) lower    & $45.6$    & $45.6^d$ & $2.559$         \\
         & Restored & (ii) centroid & $36.9$    & $36.9^d$ & $2.681$         \\
         & Restored & (ii) upper    & $34.2$    & $34.2^d$ & $2.575$         \\
         & Restored & (ii) lower    & $32.3$    & $32.3^d$ & $2.537$         \\
\midrule[0.20pt]
{\bf O5} &          &               & $66.5$    & $21.1$   & $2.990\pm0.017$ \\
{\bf B5} & J01      & (i) centroid  & $71.0$    & $22.5$   & $2.814$         \\
         & J01      & (i) upper     & $71.2$    & $21.7$   & $2.737$         \\
         & J01      & (i) lower     & $70.5$    & $21.8$   & $2.769$         \\
         & J01      & (ii) centroid & $65.0$    & $21.1$   & $2.727$         \\
         & J01      & (ii) upper    & $66.7$    & $21.9$   & $2.698$         \\
         & J01      & (ii) lower    & $68.5$    & $22.9$   & $3.200$         \\
         & Restored & (ii) centroid & $61.1$    & $24.2$   & $2.858$         \\
         & Restored & (ii) upper    & $59.3$    & $25.1$   & $2.938$         \\
         & Restored & (ii) lower    & $55.6$    & $23.3$   & $2.885$         \\
\midrule[0.20pt]
{\bf O6} &          &               & $ 96.3$   & $29.8$   & $3.460\pm0.022$ \\
{\bf B6} & J01      & (i) centroid  & $102.9$   & $31.9$   & $3.168$         \\
         & J01      & (i) upper     & $105.1$   & $33.9$   & $3.263$         \\
         & J01      & (i) lower     & $106.3$   & $35.8$   & $3.325$         \\
         & J01      & (ii) centroid & $ 90.0$   & $25.0$   & $2.852$         \\
         & J01      & (ii) upper    & $100.1$   & $33.4$   & $3.166$         \\
         & J01      & (ii) lower    & $103.1$   & $34.6$   & $4.035$         \\
         & Restored & (ii) centroid & $ 97.6$   & $36.5$   & $3.440$         \\
         & Restored & (ii) upper    & $ 95.2$   & $35.9$   & $3.414$         \\
         & Restored & (ii) lower    & $ 90.1$   & $34.5$   & $3.402$         \\
\midrule[0.20pt]
     B7  & J01      & (i) centroid  & $155.5$   & $52.6$   & $4.084$         \\
         & J01      & (i) upper     & $159.4$   & $54.3$   & $4.047$         \\
         & J01      & (i) lower     & $158.4$   & $52.1$   & $4.016$         \\
         & J01      & (ii) centroid & $128.2$   & $38.2$   & $3.441$         \\
         & J01      & (ii) upper    & $155.6$   & $55.5$   & $4.039$         \\
         & J01      & (ii) lower    & $158.5$   & $55.4$   & $4.034$         \\
         & Restored & (ii) centroid & $156.8$   & $59.2$   & $4.359$         \\
         & Restored & (ii) upper    & $154.0$   & $58.8$   & $4.393$         \\
         & Restored & (ii) lower    & $148.0$   & $57.9$   & $4.441$         \\
\bottomrule[1.0pt]
\end{tabular*}
\begin{minipage}{\linewidth}
\footnotesize
\vskip5pt
{\sc Notes}---\\
$^a$ Observed bursts and the modeled counterparts are labeled in bold texts. $^b$ The uncertainties of observational time are less than 1~s and are omitted. $^c$ The time coordinate of the O1 PRE burst is normalized to be the commencement time of epoch Oct~2002. $^d$ The observed interval between O1 and O4 bursts is taken as the recurrence time of the O4 burst. The modeled counterparts follow the same assumption. 
\end{minipage}
\end{table}
\renewcommand{\arraystretch}{1.0}

\begin{figure}[t]
\begin{center}
\includegraphics[width=\columnwidth, angle=0]{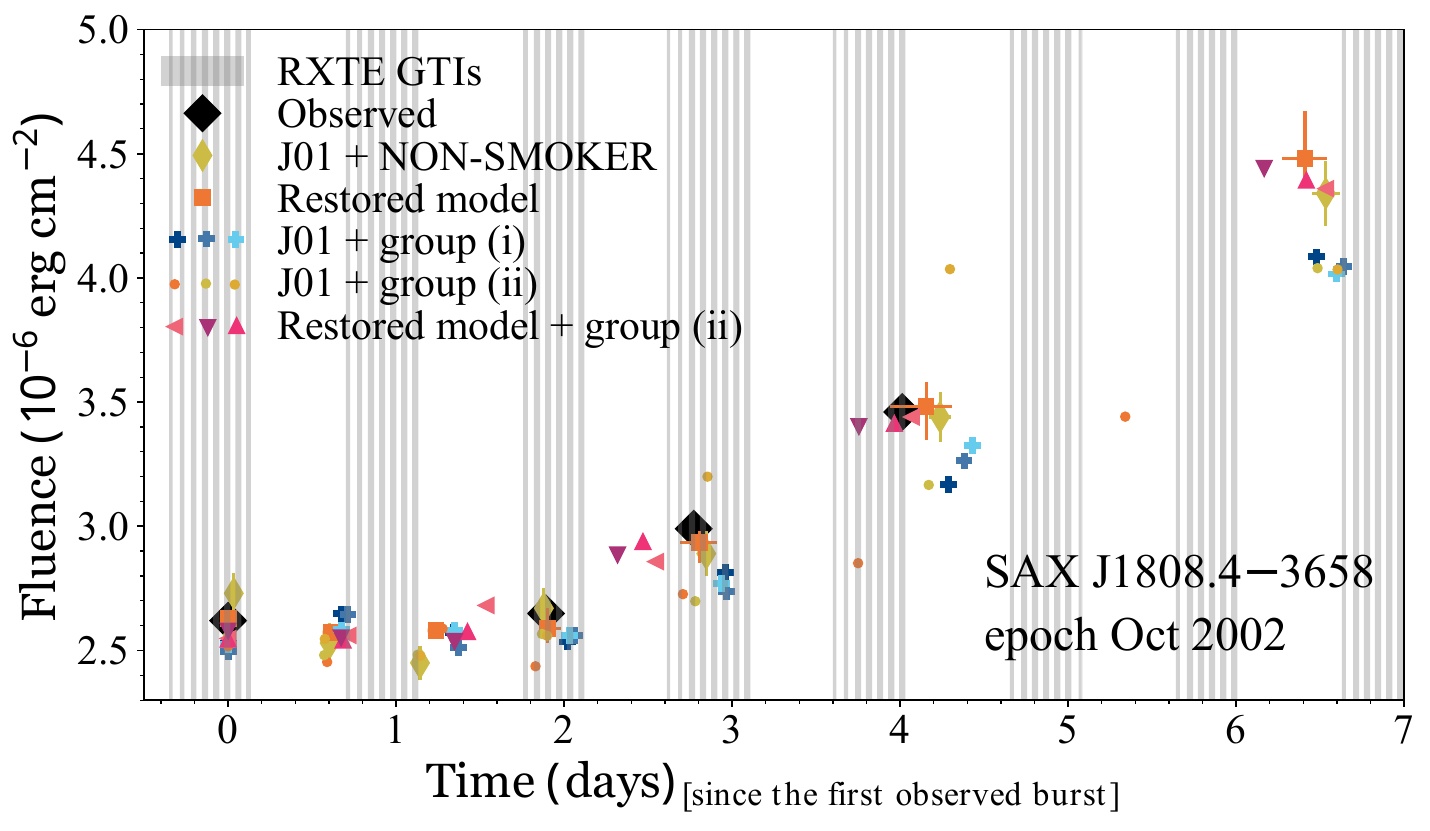}
\caption{\label{fig:PRE_fluence}{
The PRE bursts’ fluences and GTIs for the SAX~J1808.4$-$3658 burster (epoch 2002 October). The RXTE observation data (\citealt{MINBAR,Galloway2017}; black diamonds) are plotted together with the theoretical counterparts generated from the J01 PRE burst model (\citealt{Johnston2018}; yellow diamonds). The restored PRE burst model is from (\citealt{Hu2021}; orange squares), and present calculations. 
Modeled burst fluences constructed from the J01 PRE burst model using the centroid (dark blue plus signs), upper (blue plus signs), and lower (cyan plus signs) limits of \textcolor{black}{the reaction rates of group (i)}. 
Modeled burst fluences yielded from the J01 PRE burst model using the centroid (orange dots), upper (yellow dots), and lower (light yellow dots) limits of \textcolor{black}{the reaction rates of group (ii)}. 
Modeled burst fluences generated from the restored PRE burst model~\citep{Hu2021} using the centroid (red left triangles), upper (pink up triangles), and lower (purple down triangles) limits of \textcolor{black}{the reaction rates of group (ii)}. 
}}
\end{center}
\end{figure}

Four PRE bursts from the SAX~J1808.4$-$3658 X-ray source were observed in epoch 2002 October during the good time intervals (GTIs) of RXTE. These four observed (modeled) PRE bursts are labeled as O1 (B1), O4 (B4), O5 (B5), and O6 (B6), listed in Table~\ref{tab:PRE_fluence} and plotted in Fig.~\ref{fig:PRE_flux} in accord with the sequence of observation. The models also predict the possible B2 and B3 PRE bursts, which are ignited between the O1 (B1) and O4 (B4) bursts, and also the B7 PRE burst that could happen $\sim$2.5 days after the O6 burst. These predicted B2, B3, and B7 bursts are not in the GTI window. The burst fluences of the PRE bursts presented in Fig.~\ref{fig:PRE_flux} and the predicted B2, B3, and B7 bursts are plotted accordingly as well. 


The burst light-curve profiles of the B1 bursts based on the centroid, upper, and lower limits of the reaction rates of groups (i) or (ii) are rather consistent, and the burst tails are below the observed counterpart (first row subfigures in Fig.~\ref{fig:PRE_flux}); thus, the average fluence of these B1 bursts is $2.539^{+0.040}_{-0.084}\!\times\!10^{-6}$~erg$\cdot$cm$^{-2}$, which is about $0.081\!\times\!10^{-6}$~erg$\cdot$cm$^{-2}$ below the observation, $2.620\!\pm\!0.021\!\times\!10^{-6}$~erg$\cdot$cm$^{-2}$ (Fig.~\ref{fig:PRE_fluence}). The modeled PRE burst fluence of the SAX~J1808.4$-$3658 PRE burster illustrates the quality of reproduction of the burst light curves. This reproduction of the PRE burst light curve and fluence is more sensitive to the change of reaction rates than the situation of reproducing the GS~1826$-$24 periodic burst light-curve profile. 

The impact of the reaction rates of groups (i) and (ii) not only gradually manifests in the PRE burst light curves from the B4 to B6 bursts (subfigures of the B4, B5, and B6 bursts in the left and middle columns of Fig.~\ref{fig:PRE_flux}) but also becomes evident in the recurrence times corresponding to the B4, B5, and B6 bursts (Table~\ref{tab:PRE_fluence} and the colored plus signs and dots in Fig.~\ref{fig:PRE_fluence}).

For the reaction rates of group (i), although the present $^{63}$Ga(p,$\gamma$)$^{64}$Ge is around a factor of 2-3 higher than the recommended \emph{ths}8 rate, when the enhanced reaction flow reaches the $^{64}$Ge(p,$\gamma$)$^{65}$As(p,$\gamma$)$^{66}$Se branch, the new $^{64}$Ge(p,$\gamma$)$^{65}$As reaction rate of 2 orders of magnitude lower than the recommended \emph{ths}8 rate reduces the influx of the reaction flow going through this branch and thus weakens the GeAs cycle. Such a reduction keeps the reaction flow in the cycles of nuclei lighter than the Ge isotopes, e.g., the NiCu and ZnGa cycles; allows a slow pace of hydrogen burning; reduces fluences; and delays the onsets of the B4, B5, B6, and B7 bursts (Table~\ref{tab:PRE_fluence} and the colored plus signs in Fig.~\ref{fig:PRE_fluence}).


For the reaction rates of group (ii), the updated $^{22}$Mg($\alpha$,p)$^{25}$Al reaction rate is more than 1 order of magnitude lower than the recommended \emph{ths}8 rate. The updated rate restricts the $^{22}$Mg($\alpha$,p) branch and permits the reaction flow more siphoned out to the $^{22}$Mg(p,$\gamma$)$^{23}$Al(p,$\gamma$)$^{24}$Si($\alpha$,p) branch, enriching more proton-rich nuclei near the drip line past the \emph{sd}-shell nuclei and further enhancing hydrogen burning. Despite the greater enrichment of proton-rich nuclei around the drip line, the reaction flow and hydrogen burning are still limited by the weak GeAs cycle with the new $^{64}$Ge(p,$\gamma$)$^{65}$As and $^{65}$As(p,$\gamma$)$^{66}$Se reaction rates. Such a limit removes the ignition of an extra onset of the PRE burst between the B1 and B4 bursts as found by \citet{Hu2021}. For the models using the centroid or upper limit of the reaction rates of group (ii), the accreted hydrogen is depleted faster than the models using the lower limit of the reaction rates of group (ii) or the models using the reaction rates of group (i) (without the updated $^{22}$Mg($\alpha$,p)$^{25}$Al reaction rate). The early depletion of the accreted hydrogen somehow causes the decrement of the burst tail (subfigures of the centroid and upper limits of the B4, B5, and B6 bursts in the second column of Fig.~\ref{fig:PRE_flux}), thus causing the decrease of the fluences of the B4, B5, and B6 bursts (orange and yellow dots in Fig.~\ref{fig:PRE_fluence}).

Adopting the restored model ($X(\mathrm{He})\!=\!0.567\!\pm\!0.003$ for the base of the accreted envelope) by \citet{Hu2021} for studying the influence of the reaction rates of group (ii), we find that PRE burst light curves for the O1, O4, O5, and O6 bursts are closer to the observed counterparts compared to the B1, B4, B5, and B5 bursts in the first and second columns of Fig.~\ref{fig:PRE_flux} (see subfigures of the right column of Fig.~\ref{fig:PRE_flux}), with merely a $2.4^{+2.0}_{-1.8}$~\% difference in comparing the observed and modeled fluences. The burst tails of B4 and B5 are lower than the observed ones, still indicating an early depletion of hydrogen and causing a $3.0^{+1.4}_{-1.8}$~\% difference in fluences compared to the observed counterparts. As the restored model implements $6$~\% more of He mass fraction for the base of the accreted envelope, such redefined enrichment of He shortens the time for the pure helium layer below the hydrogen-burning shell to reach the ignition threshold and ignites the helium-shell flash early, causing the onset times of B4, B5, and B6 to be overall $12.9^{+15.9}_{-11.8}$~\% earlier than the observed counterparts (colored triangles in Fig.~\ref{fig:PRE_fluence}) and removing the B3 burst in between the B1 and B4 bursts. After ignition in the pure helium layer, the enriched helium in the base also offsets the flow to the $^{22}$Mg(p,$\gamma$)$^{23}$Al(p,$\gamma$)$^{24}$Si($\alpha$,p) branch, allowing slower hydrogen burning for the onsets and prolonging the depletion of hydrogen at the burst tail stage.

Overall, the SAX~J1808.4$-$3658 PRE burst is more sensitive than the GS~1826$-$24 periodic burst to the new reaction rates defined in the reaction rates of groups (i) and (ii) in terms of reproducing the observed PRE burst light curves and fluences due to the depletion rate of the accreted hydrogen and the GeAs cycle that resides in the rp-process path in the region around nuclear masses $A\!=\!60$-$64$. The new reaction rates of group (ii) suggest a necessary fine-tuning of the restored PRE burst model that we leave for future work. Future direct measurements of $^{22}$Mg($\alpha$,p) at the XRB Gamow energy are crucial to constrain the restored and future developed PRE models. 
\section{Summary and Conclusion}
\label{sec:summary}


We study the influence of new $^{63}$Ga(p,$\gamma$)$^{64}$Ge and $^{64}$Ge(p,$\gamma$)$^{65}$As reaction rates, and the experimentally determined $^{22}$Mg($\alpha$,p)$^{25}$Al, on the periodic bursts of the GS~1826$-$24 clocked burster (epoch 2007 March) and the PRE bursts of the SAX~J1808.4$-$3658 burster (epoch 2002 October). These reaction rates are grouped as 
(i) $^{63}$Ga(p,$\gamma$)$^{64}$Ge and $^{64}$Ge(p,$\gamma$)$^{65}$As and 
(ii) $^{63}$Ga(p,$\gamma$)$^{64}$Ge, $^{64}$Ge(p,$\gamma$)$^{65}$As, and $^{22}$Mg($\alpha$,p)$^{25}$Al of \citet{Hu2021}. 
For both group (i) and (ii), we further use the recently updated reaction rates, i.e., $^{55}$Ni(p,$\gamma$)$^{56}$Cu \citep{Valverde2019}, $^{56}$Ni(p,$\gamma$)$^{57}$Cu \citep{Kahl2019}, $^{65}$As(p,$\gamma$)$^{66}$Se \citep{Lam2022a}, and $^{57}$Cu(p,$\gamma$)$^{58}$Zn \citep{Lam2022b}.

\textcolor{black}{The construction of the GS~1826$-$24 clocked burst models (epoch 2007 March) based on JINA REACLIB v2.2, or the reaction rates of group (i) or (ii) define the baseline, \emph{Present}$^\dag$, or \emph{Present}$^\ddag$ models, respectively. 
We follow the procedure used by \citet{Hu2021} and \citet{Lam2022a,Lam2022b} to construct these models on an equal footing, implementing the same astrophysical configurations and aiming to closely reproduce the observed burst light-curve profile, fluence, burst peak, and recurrence time. The capability of the baseline, \emph{Present}$^\dag$, and \emph{Present}$^\ddag$ models is exhibited by reproducing the observed fluence ($1.18\!\pm0.04\!\times\!10^{-6}$~erg$\cdot$cm$^{-2}$), recurrence time ($\Delta t_\mathrm{rec}\!=\!3.530\pm0.004$~hr), and burst peak ($28.4\pm0.4\!\times\!10^{-9}$~erg$\cdot$cm$^{-2}$s$^{-1}$) with low uncertainties. Basically, the modeled observables agree well with the observed counterparts (Fig.~\ref{fig:periodic_observables}), and the average deviation between the modeled and observed light curve is limited to only 
\textcolor{black}{$0.55\!\times\!10^{-9}$}erg$\cdot$cm$^{-2}$s$^{-1}$ for the baseline, 
\textcolor{black}{$0.52\!\times\!10^{-9}$}erg$\cdot$cm$^{-2}$s$^{-1}$ for \emph{Present}$^\dag$, and
\textcolor{black}{$0.45\!\times\!10^{-9}$}erg$\cdot$cm$^{-2}$s$^{-1}$ for \emph{Present}$^\ddag$. 
}


Comparing both modeled burst light-curve profiles and burst fluences of the baseline and \emph{Present}$^\dag$ (Figs.~\ref{fig:periodic_flux}), we find that \textcolor{black}{the periodic burst light curve of the GS~1826$-$24 clocked burster (epoch 2007 March) is, in fact, not sensitive to the new forward and reverse $^{64}$Ge(p,$\gamma$)$^{65}$As and $^{65}$As(p,$\gamma$)$^{66}$Se reaction rates with the newly determined proton thresholds of $^{65}$As and $^{66}$Se.} Our finding is consistent with the sensitivity study of \citet{Cyburt2016} \textcolor{black}{but contradicts the result of \citet{Zhou2023}, emphasizing the importance of (i) constructing the initial baseline and all comparison models in equal footing and (ii) impartially fitting the modeled burst light curves to the entire burst time span of the averaged-observed light-curve profile for an objective study of the impact of the newly deduced reaction rates. As both important steps were not considered in the models used by \citet{Zhou2023}, the impact of the newly determined proton thresholds of $^{65}$As and $^{66}$Se on the GS~1826$-$24 periodic burst light curves \citet{Zhou2023} is somehow subjectively overestimated.} Also, the uncertainty induced from the pronounced mismatch propagates to the neutron star compactness and burst ash composition deduced from their models. This indicates that the reliability of any inferences deduced from reproducing a set of XRB observables, for instance but not limited \textcolor{black}{to the} burst light-curve profile, burst peak, fluence, and recurrence time, depends on the qualitative and quantitative reconciliation of the modeled observables and the observed counterparts. This prerequisite underlines the improvement we shall \textcolor{black}{consider for} constructing new XRB models, which are highly desired within the community to deduce the neutron star compactness and obtain a reliable burst ash composition for neutron star crust cooling models. 




A more rigorous approach, the multiepoch match performed by \citet{Johnston2020}, is proposed (\citealt{Galloway_private_2022}, private communication). Nevertheless, we note that the finding of the impact of the new reaction rates on the GS~1826$-$24 periodic burst light-curve profile and the recurrence time of this work (epoch 2007 March) remains valid on epochs 1998 June and 2000 September. For instance, the present study on epoch 2007 March for the experimentally determined $^{22}$Mg($\alpha$,p)$^{25}$Al reaction rate has the same impact as the finding on epoch 1998 June by \citet{Hu2021} and \citet{Lam2022a,Lam2022b}. Moreover, tremendous work to construct new reaction rates for key reactions is required to further stringently constrain the GS~1826$-$24 periodic burst models for the multiepoch match. The success of a new multiepoch match based on updating all key reactions will further constrain the important astrophysical configurations and neutron star properties, and, importantly, it marks the milestones of reconstructing the GS~1826$-$24 clocked bursts that have been observed since its discovery. We leave this nontrivial work for exploration in the near future, and the results of this work will be published elsewhere.

We find that the SAX~J1808.4$-$3658 PRE burster is more sensitive to the new forward and reverse $^{64}$Ge(p,$\gamma$)$^{65}$As and $^{65}$As(p,$\gamma$)$^{66}$Se reaction rates with the newly determined proton thresholds of $^{65}$As and $^{66}$Se because (i) the recurrence times are nonconstant and the burst profiles are irregular, (ii) the accreted hydrogen is depleted much faster than the case of GS~1826$-$24 periodic bursts, causing the rp-process path is limited up to the region of nuclear masses $A\!=\!60$-$64$, and (iii) the GeAs cycle resides in that region. 
The impact of the new $^{64}$Ge(p,$\gamma$)$^{65}$As and $^{65}$As(p,$\gamma$)$^{66}$Se reaction rates on the PRE burst light curves, burst fluences, and recurrence times of SAX~J1808.4$-$3658 PRE burster (epoch 2002 October) is shown in Figs.~\ref{fig:PRE_flux}~and~\ref{fig:PRE_fluence}. Hence, using the new reaction rates of group (ii) necessitates a further fine-tuning for the restored PRE burst model to closely reproduce the burst light curves, fluences, and recurrence times. The finding of the present work agrees with the preliminary result of \citet{Lam2022c} that the PRE bursts of SAX~J1808.4$-$3658 (epoch 2002 October) are more sensitive to variations of thermonuclear reaction rates than the periodic bursts of the GS~1826$-$24 burster. The results of sensitivity study of SAX J1808.4--3658 PRE burst model will be published elsewhere (Y. H. Lam et al. 2025, in preparation).

\hfill\break
\noindent
{\bf\large Acknowledgments}\\
\indent 
We deeply appreciate the anonymous Referee for carefully reviewing our manuscript with useful suggestions and comments. We are very grateful to Jun Hu (Institute of Modern Physics, Chinese Academy of Sciences, China) and Michael S. Smith (Oak Ridge National Laboratory, US) for fruitful discussion on the experimental aspect of $^{22}$Mg($\alpha$,p)$^{25}$Al reaction rates and to Duncan K. Galloway (Monash University, Australia) for understanding the multiepoch match. 
This work was supported by the Strategic Priority Research Program of the Chinese Academy of Sciences (CAS; grant Nos. XDB34020204 and XDB34020100), the Ministry of Science and Technology of the People's Republic of China (National Key Research and Development Program of China; No. 2021YFA1601500), and the National Natural Science Foundation of China (Nos. 12375146, 11775277, 11961141004). 
We greatly appreciate the computing resources, i.e., PHYS\_T3 cluster, Distributed Cloud resources (FDR5 cluster), provided by the Institute of Physics and Academia Sinica Grid-computing Center of Academia Sinica (ASGC; grant No. AS-CFII-112-103), Taiwan. 
Some of the numerical computations of XRBs in this work were carried out at the Yukawa Institute Computer Facility, i.e., Yukawa-21 at the Yukawa Institute of Theoretical Physics (YITP) of Kyoto University, and at the general-purpose PC cluster of the Center for Computational Astrophysics (CfCA) of National Astronomical Observatory of Japan.
Y.H.L. gratefully acknowledges the financial support from the Science Foundation of Zhejiang Sci-Tech University (No. 25062123-Y) and the Chinese Academy of Sciences President's International Fellowship Initiative (No. 2019FYM0002), the graceful hospitality from Shigeru Kubono and Daisuke Suzuki (RIKEN, the University of Tokyo, Japan) for visiting RIKEN in 2024 February, the financial support from Katsuhisa Nishio (Japan Atomic Energy Agency, Japan) for participating in the \href{https://asrc.jaea.go.jp/soshiki/gr/HENS-gr/NAPS2024/}{ASRC Workshop}, and the strong support from Nobuya Nishimura (CNS, the University of Tokyo, Japan) and Akira Dohi (RIKEN, Japan) for participating in the \href{https://indico.yukawa.kyoto-u.ac.jp/event/46/}{Nucleosynthesis and Evolution of Neutron Stars Workshop} and for visiting RIKEN in 2025 January-February. A.H. is supported by the Australian Research Council Centre of Excellence for Gravitational Wave Discovery (OzGrav; No. CE170100004) and for All Sky Astrophysics in 3 Dimensions (ASTRO 3D; No. CE170100013) and is also supported in part by the US National Science Foundation under grant No. PHY-1430152 (JINA Center for the Evolution of the Elements, JINA-CEE). 
Z.X.L. is supported by the Talent Support Project of Huizhou (Innovation Team 2023; No. 230806198702053). H.Y. is supported by JSPS KAKENHI (No. 23K25877, No. 18H01218, No. 19K03883, and No. 23H01181) from the Ministry of Education, Culture, Sports, Science and Technology (MEXT) of Japan. 





\newpage
{

}



\end{document}